\documentclass[aps,prl,twocolumn,showpacs,10pt,superscriptaddress,preprintnumbers,nofootinbib]{revtex4-1}
\usepackage{epsfig,amssymb,amsmath,psfrag,epstopdf,color}
\pdfoutput=1

\usepackage{graphicx}
\usepackage{subcaption}
\usepackage{ragged2e}
\DeclareCaptionJustification{justified}{\justifying}
\usepackage[normalem]{ulem}
\usepackage{paralist}
\usepackage{hyperref}
\usepackage[size=footnotesize]{todonotes}
\allowdisplaybreaks
 
\usepackage{tikz}
\usepackage{mathrsfs}
\usepackage{hyperref}
\usepackage{multirow}
\usepackage{scalerel}
\usepackage{mathtools}
\usepackage{textcomp}
\usepackage[all]{xy}
\usepackage{dsfont}
\usepackage{amsthm}
\usepackage{blkarray}
\usepackage{bm}
\usepackage{todonotes}
\usepackage{tikz-cd}
 
\usepackage{graphicx}
\usepackage{amsmath,amssymb,mathtools} 
\usepackage{xspace}
\usepackage{wrapfig}
\usepackage{slashed}
\usepackage[normalem]{ulem}
\usepackage[utf8]{inputenc}
\usepackage[T1]{fontenc}
\usepackage{mathtools}
\usepackage{stmaryrd}
\usepackage{appendix}

\usepackage{scrextend}

\usepackage{tabularx}
\usepackage{tensor}

\newcommand{\eq}[1]{Eq.~\eqref{eq:#1}}

\newcommand{\tab}[1]{Tab.~\ref{tab:#1}}

\DeclareRobustCommand{\fig}[1]{Fig.~\ref{fig:#1}}

\usepackage{marginnote}

\definecolor{darkgreen}{rgb}{0.0, 0.4, 0.0}

\newcommand{\df}{\mathrm{d}}

\newcommand{\ra}{\rightarrow}

\newcommand{\as}{\alpha_s}

\newcommand{\nn}{\nonumber}

\newcommand{\zcut}{z_{\rm cut}}

\newcommand{\Ok}{\Omega_{1\kappa}^{\figeight}}
\newcommand{\Uka}{\Upsilon_{1,0\kappa}^{\bndry}}
\newcommand{\Ukb}{\Upsilon_{1,1\kappa}^{\bndry}}

\newcommand{\Pythiaxx}{\texttt{Pythia\xspace8.306}\xspace}
\newcommand{\Pythia}{\texttt{Pythia}\xspace}
\newcommand{\Vincia}{\texttt{Vincia}\xspace}
\newcommand{\Vinciaxx}{\texttt{Vincia\xspace2.3}\xspace}
\newcommand{\Herwig}{\texttt{Herwig}\xspace}
\newcommand{\Herwigxx}{\texttt{Herwig\xspace7.2.3}\xspace}

\newcommand{\letter}{\emph{Letter}\xspace}

\def\df{\textrm{d}}
\def\nn{\nonumber}

\def\LQCD{\Lambda_{\rm QCD}}

\def\bndry{\varocircle}
\def\figeight{{\circ\!\!\circ}}

\newcommand{\ee}{e^+e^-}

\allowdisplaybreaks[2]
\newcommand{\Oq}{\Omega_{1q}^{\figeight}}
\newcommand{\Uqa}{\Upsilon_{1,0q}^{\bndry}}
\newcommand{\Uqb}{\Upsilon_{1,1q}^{\bndry}}

\newcommand{\Og}{\Omega_{1g}^{\figeight}}
\newcommand{\Uga}{\Upsilon_{1,0g}^{\bndry}}
\newcommand{\Ugb}{\Upsilon_{1,1g}^{\bndry}}

\DeclareRobustCommand{\Reff}[1]{Ref.~\cite{#1}}

\DeclareRobustCommand{\eq}[1]{Eq.~(\ref{eq:#1})}


\begin{document}
\preprint{
    \vbox{\hbox{MIT--CTP 5475}\hbox{DESY--22--159}}
}

\title{Field-Theoretic Analysis of Hadronization Using Soft Drop Jet Mass} 

\author{Anna Ferdinand}%
\email{aef54@cam.ac.uk}
\affiliation{University of Manchester, School of Physics and Astronomy, Manchester, M13 9PL, United Kingdom}
\affiliation{DAMTP, University of Cambridge, Wilberforce Road, Cambridge, CB3 0WA, United Kingdom}
\author{Kyle Lee}%
\email{kylel@mit.edu}
\affiliation{Nuclear Science Division, Lawrence Berkeley National Laboratory, Berkeley, California 94720, USA}
\affiliation{Center for Theoretical Physics, Massachusetts Institute of Technology, Cambridge, MA 02139, USA}
\author{Aditya Pathak}%
\email{aditya.pathak@desy.de}
\affiliation{University of Manchester, School of Physics and Astronomy, Manchester, M13 9PL, United Kingdom}
\affiliation{Deutsches Elektronen-Synchrotron DESY, Notkestr. 85, 22607 Hamburg, Germany}

\begin{abstract}
One of the greatest challenges in quantum chromodynamics is understanding the hadronization mechanism, which is also crucial for carrying out precision physics with jet substructure. In this \emph{Letter}, we combine recent advancements in our understanding of the field theory-based nonperturbative structure of the soft drop jet mass with precise perturbative calculations of its multi-differential variants at next-to-next-to-leading logarithmic (NNLL) accuracy. This enables a systematic study of its hadronization power corrections in a completely model-independent way. We calibrate and test hadronization models and their interplay with parton showers by comparing our universality predictions with various event generators for quark and gluon initiated jets in both lepton-lepton and hadron-hadron collisions. We find that hadronization models perform better for quark jets relative to gluon jets. Our results provide the necessary toolkit for precision studies with the soft drop jet mass motivating future analyses using real world collider data. The nontrivial constraints derived in our framework are useful for improving the modeling of hadronization and its interface with parton showers in next generation event generators.
\end{abstract}

\maketitle

The study of jets and their substructure has become a very active program at high energy particle colliders in the last decade~\cite{Larkoski:2017jix,Kogler:2018hem}. A key development has been the use of jet grooming techniques~\cite{Ellis:2009me,Cacciari:2014gra,Krohn:2009th,Dasgupta:2013ihk,Larkoski:2014wba,Butterworth:2008iy,Frye:2017yrw,Dreyer:2018tjj} that allow for theoretical control by eliminating contamination from the wide-angle soft radiation from the underlying event and pile-up, and by reducing hadronization effects. In particular, the soft drop (SD) grooming~\cite{Larkoski:2014wba,Butterworth:2008iy,Dasgupta:2013ihk} has received the most widespread attention, inspiring many theoretical calculations both for jets in vacuum~\cite{Larkoski:2017cqq,Larkoski:2017iuy,Baron:2018nfz,Kang:2018vgn,Makris:2018npl,Kardos:2018kth,Napoletano:2018ohv,Lee:2019lge,Gutierrez-Reyes:2019msa,Kardos:2019iwa,Mehtar-Tani:2019rrk,Kardos:2020ppl,Larkoski:2020wgx,Lifson:2020gua,Caucal:2021cfb,Chien:2016led,Milhano:2017nzm,Chang:2017gkt,Li:2017wwc,Chien:2018dfn,Sirimanna:2019bgl,Caucal:2019uvr,Ringer:2019rfk,Cal:2021fla,Larkoski:2017bvj,Chien:2019osu,Cal:2020flh,Stewart:2022ari} and in medium~\cite{KunnawalkamElayavalli:2017hxo,Andrews:2018jcm,Mehtar-Tani:2016aco,Casalderrey-Solana:2019ubu,Brewer:2021hmh}, as well as several experimental analyses~\cite{Aaboud:2017qwh,Sirunyan:2017bsd,Kauder:2017cvz,Sirunyan:2018xdh,Acharya:2019djg,ATLAS:2019sol,Aad:2019vyi,Adam:2020kug,Aad:2020zcn,Sirunyan:2017bsd,Sirunyan:2018gct, Acharya:2019djg,Kauder:2017cvz}.
Among various groomed observables, the SD jet mass is by far the most extensively studied within both theoretical~\cite{Dasgupta:2013ihk,Dasgupta:2013via,Larkoski:2014pca,Dasgupta:2015yua,Marzani:2017mva,Chay:2018pvp,Kang:2018jwa,Marzani:2017kqd} and experimental communities~\cite{ALICE:2017nij,ATLAS:2017zda,ATLAS:2018jsv,ATLAS:2019dty,ATLAS:2019mgf,CMS:2017tdn,CMS:2018vzn,CMS:2018fof,ALICE:2021njq,STAR:2021lvw,HKlest2022}, and has been explored for a variety of phenomenological applications, such as quantifying medium modification~\cite{CMS:2018fof,ALICE:2017nij,KunnawalkamElayavalli:2017hxo}, and precision top quark mass~\cite{Hoang:2017kmk,Bachu:2020nqn,ATLAS:2021urs} and strong coupling constant~\cite{Marzani:2019evv,Hannesdottir:2022rsl}  measurements.

\begin{figure}[t]
	\centering
	\includegraphics[width=0.45\textwidth]{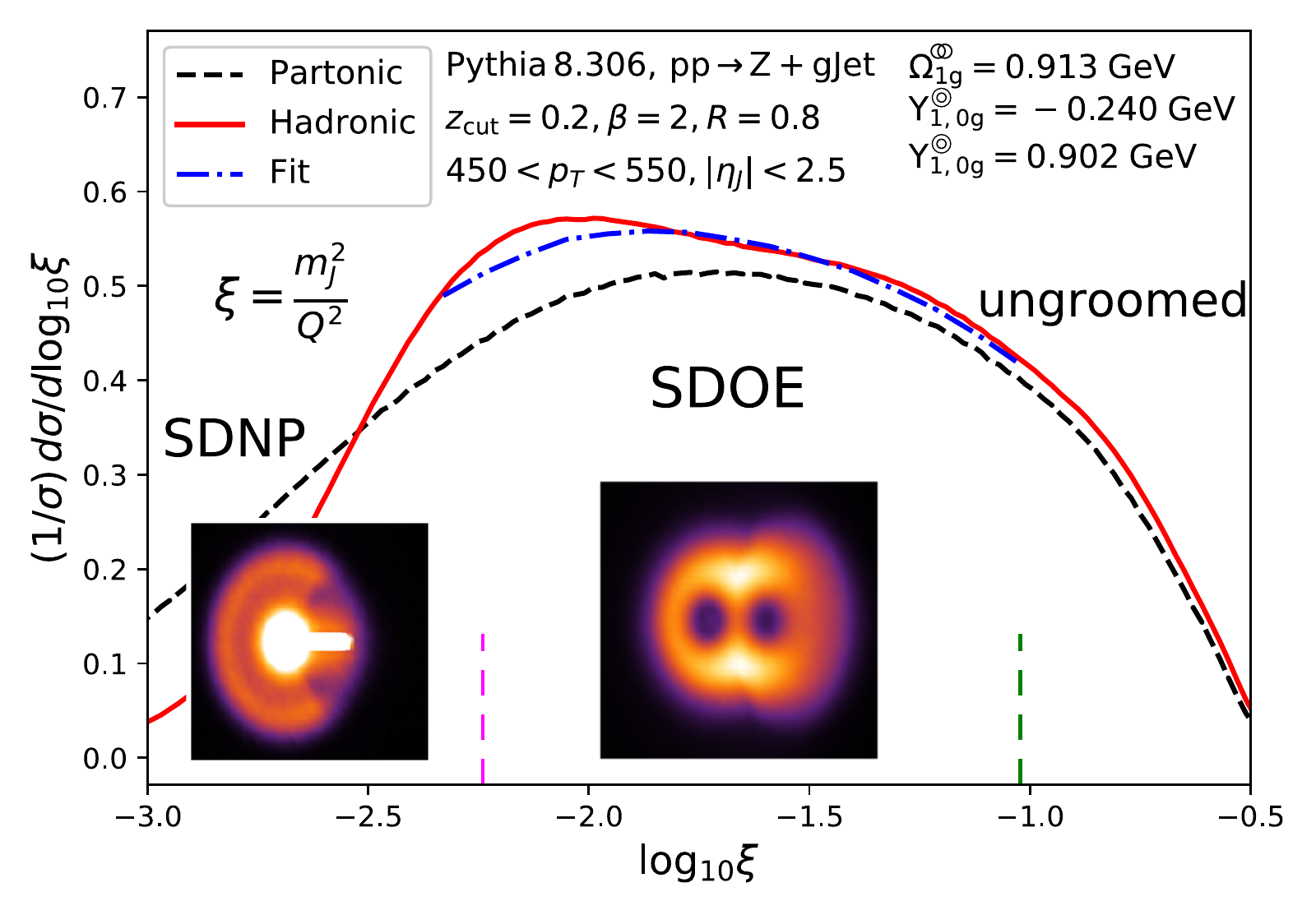}
	\vspace{-5pt}
	
	\caption{An example of fit for nonperturbative parameters in \Pythiaxx simulation of groomed jet mass.   The insets show distribution of low energy particles as heat maps around the soft drop stopping subjets in the transverse plane. 
	}	\label{fig:PythiaFit}
\end{figure}

Theoretically, the SD jet mass is the most precisely studied groomed jet observable, with predictions available at next-to-next-to-next-to-leading logarithmic accuracy (N$^3$LL) matched to next-to-next-to-leading order (NNLO) predictions for dijets at $\ee$ collisions~\cite{Kardos:2020gty}, and next-to-next-to-leading-logarithmic (NNLL) accuracy~\cite{Hannesdottir:2022rsl} for jets at the LHC. At this level of precision, the hadronization power corrections become comparable in size to perturbative accuracy, and cannot be accounted for using hadronization models~\cite{Sjostrand:2007gs,Gleisberg:2008ta,Bahr:2008pv} that are tuned to lower precision parton showers. 
In the recent years, there has been significant progress in understanding these hadronization effects in the SD jet mass~\cite{Hoang:2019ceu,Pathak:2020iue,AP} using a field theory-based formalism~\cite{Lee:2006fn,Korchemsky:1994is,Korchemsky:1997sy,Korchemsky:1999kt,Belitsky:2001ij,Dasgupta:2007wa,Stewart:2014nna}, which allows for a model-independent description of nonperturbative (NP) power corrections for precision phenomenology. Furthermore, this formalism imposes powerful constraints on jet-flavor, kinematics and grooming parameter dependence of these NP corrections. 
Hence, by comparing these predictions with event generators, we now have a unique opportunity to carry out a nontrivial characterization of these hadronization models and their interplay with parton showers, which is often difficult to interpret and test. In this \letter, using the state-of-the-art theoretical advancements in our understanding of the SD jet mass, we achieve a systematic and complete field theory-based study of hadronization effects and test our predictions with multiple event generators.

\emph{Hadronization corrections to groomed jet mass.}--- Compared to the ungroomed jet mass, the SD jet mass exhibits a much larger region of applicability for perturbation theory. This region is referred to as the soft drop operator expansion (SDOE) region, which is defined below and shown in \fig{PythiaFit} between the vertical lines. Here, hadronization effects can be studied using factorization in a systematic expansion. 

Using soft collinear effective theory (SCET)~\cite{Bauer:2000ew,Bauer:2000yr,Bauer:2001ct,Bauer:2001yt}, in \Reff{Hoang:2019ceu} the leading hadronization corrections in the SDOE region were shown to depend on three ${\cal O}(\Lambda_{\rm QCD})$ NP universal constants $\{\Ok, \Uka, \Ukb \}$, which solely depend on
the parton $\kappa = q,g$ initiating the jet, and are completely independent of the jet kinematics, such as the jet $p_T$ (or $E_J$), rapidity $\eta_J$, radius $R$, and the SD parameters~\cite{Larkoski:2014wba}, the energy cut $\zcut$ and the angular modulation parameter $\beta$, such that
\begin{align}\label{eq:NP}
    \frac{1}{\sigma_\kappa} \frac{\df \sigma_\kappa}{\df m_J^2} 
	&= \frac{1}{\hat \sigma_\kappa} \frac{\df \hat\sigma_\kappa }{\df m_J^2} - Q \Ok \frac{\df }{\df m_J^2} \frac{1}{\hat \sigma_\kappa} \frac{\df \hat\sigma_\kappa^{\figeight} }{\df m_J^2} \\
	&	\quad + \frac{\Uka + \beta \Ukb}{Q}
	\frac{1}{\hat \sigma_\kappa} \frac{\df \hat\sigma_\kappa^{\bndry} }{\df m_J^2} + \cdots \,,\nn
\end{align}
Here $\df \sigma_\kappa$ and  $\df\hat \sigma_\kappa$, respectively, refer to hadron and parton level groomed jet mass cross sections for flavor $\kappa$ and $Q$ characterizing the the hard scale of the jet. The weights $\df \hat \sigma_\kappa^{\figeight,\bndry}$ are perturbatively calculable. We note that, in contrast with analytical hadronization models employed in previous work~\cite{Dasgupta:2007wa,Dasgupta:2013ihk,Marzani:2017kqd,Marzani:2019evv}, \eq{NP} is a model-independent statement and includes hadron mass effects.

In the SDOE region, the leading hadronization corrections are driven by a two-pronged dipole, which consists of an energetic collinear subjet at the core of the jet and a collinear-soft (c-soft) subjet that is responsible for stopping the grooming algorithm. The corrections represented by the ellipsis `$\ldots$' in \eq{NP} involve higher power corrections of $\LQCD$ and corrections from configurations that distort the two-pronged catchment area. The latter correction is a next-to-leading-logarithmic effect, and therefore \eq{NP} can also be seen as a factorization of NP effects at leading-logarithmic accuracy, where the strong ordering of angles ensures the two-pronged geometry. As the jet mass decreases, we enter the soft drop non-perturbative (SDNP) region, where the c-soft mode becomes nonperturbative and correspondingly the nonperturbative effects are of ${\cal O}(1)$. The transition between these two regions is clearly visible in \fig{PythiaFit}, where the insets show the distribution of low-energy NP particles in the transverse plane of the jet~\cite{Hoang:2019ceu}.

The statement of NP factorization in \eq{NP} presents us with a singular opportunity to probe hadronization in jets in a rich setting. As can be seen from \eq{NP}, the consistency of the formalism requires that the three constants be sufficient to describe data measured from high energy colliders over a wide range of energies. The highly constraining structure given by \eq{NP} (constants being of $\mathcal{O}(\Lambda_{\rm QCD})$, having a $\beta$ proportional coefficient $\Ukb$, $\zcut$-independence, etc.) 
makes this far from a trivial feat and hence useful for calibrating hadronization models. In this work, we demonstrate how the universality structure strongly constrains the NP parameters, allowing them to be accurately determined by considering various combinations of soft drop and kinematic parameters. This, for example, improves the prospects for measuring the strong coupling constant $\as$ at the LHC.

\begin{figure}[t]
	\centering
	\includegraphics[width=0.48\textwidth]{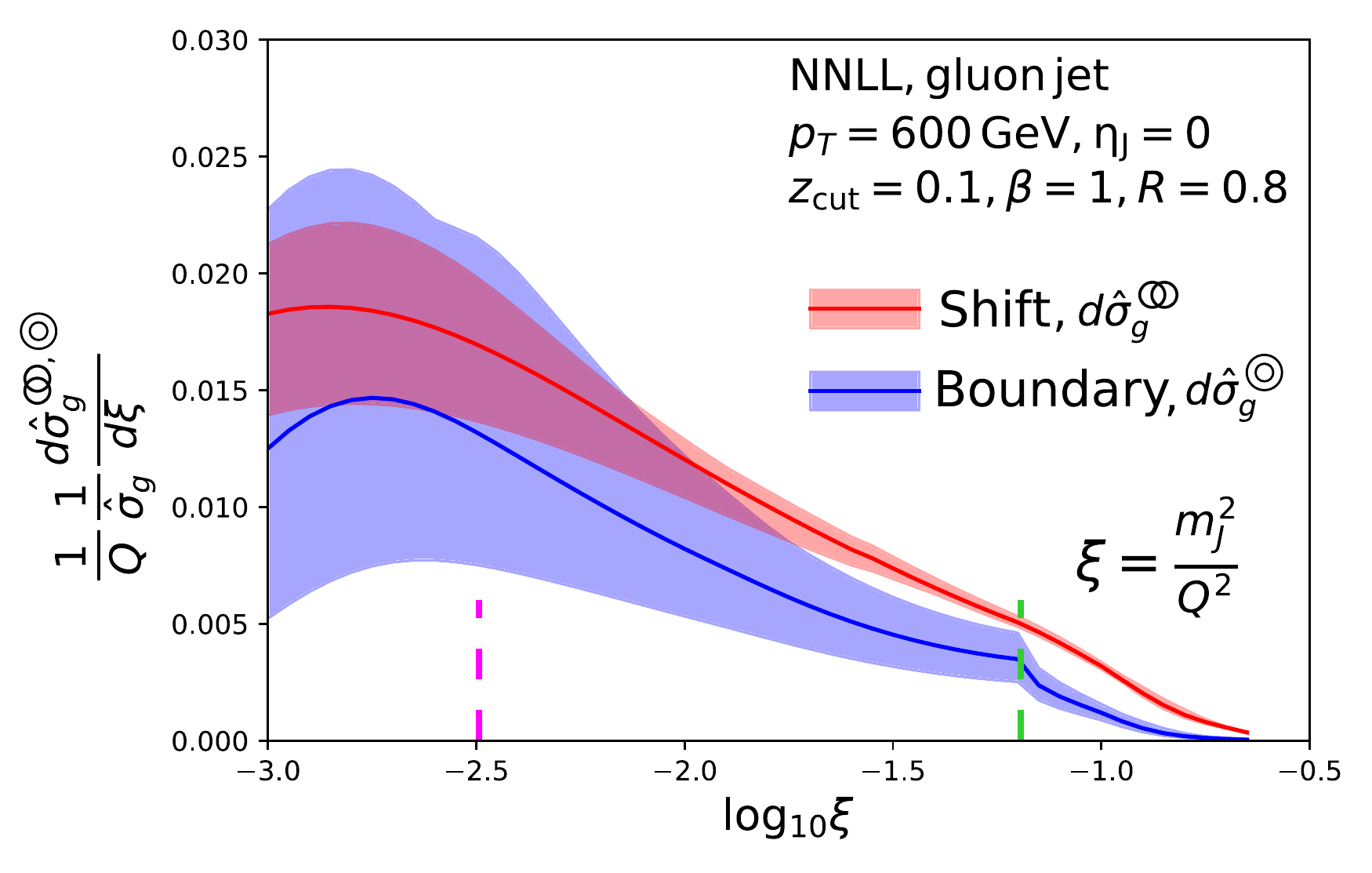}
	\caption{NNLL results for perturbative weights in \eq{NP} of hadronization corrections (shown here for gluon jets). Bands denote perturbative uncertainty and vertical lines the extent of the fit region (see \eq{xiSDOE}). The factor of $1/Q$ is included to illustrate the size of hadronization corrections.}	\label{fig:C12Sig}
\end{figure}

\emph{Calculation of perturbative weights.}---
Characterizing the two-pronged configuration of the collinear and the c-soft subjet in the SDOE region requires auxiliary measurements of the groomed jet radius $R_g$ and soft subjet energy fraction $z_g$~\cite{Cal:2021fla,Larkoski:2017bvj,Kang:2019prh,Larkoski:2014wba}, which after marginalizing give~\cite{Pathak:2020iue}
\begin{align}\label{eq:pert}
    \frac{1}{\hat \sigma_\kappa} \frac{\df \hat\sigma_\kappa^\figeight }{\df m_J^2} &\equiv \int \df r_g \: r_g \frac{1}{\hat \sigma_\kappa} \frac{\df^2 \hat \sigma_\kappa}{\df m_J^2 \df r_g} \, , \\
    \frac{1}{\hat \sigma_\kappa} \frac{\df \hat\sigma_\kappa^\bndry }{\df m_J^2}
    &\equiv
    \int \frac{\df r_g  \df z_g \: \delta\big(z_g  - \zcut r_g^\beta\big)}{r_g} \frac{1}{\hat \sigma_\kappa} \frac{ \df^3 \hat\sigma_\kappa  }{\df m_J^2 \df r_g \df z_g} \, .\nn
\end{align}
As the NP constants in \eq{NP} are independent of the jet kinematics and grooming parameters, all these dependencies are encapsulated by $\df \hat \sigma_\kappa^{\figeight,\bndry}$. The appearance of $r_g = R_g/R$ in \eq{pert} is analogous to how jet radius $R$ appears in hadronization corrections for the ungroomed jet mass in the tail and for the jet $p_T$~\cite{Dasgupta:2007wa,Stewart:2014nna}:
\begin{align}\label{eq:NPUngroomed}
 	m_{J,\rm no\,sd}^2 \!=\!  \hat m_{J,\rm no\,sd}^2 \!+\! p_TR\,\Omega_{1\kappa} \, , \quad 
	 p_T = \hat p_T +  \frac{1}{R}\Upsilon_{1\kappa}   \, , 
\end{align}
where $\Omega_{1\kappa}, \Upsilon_{1\kappa} \sim \LQCD$ are NP parameters and hatted variables are parton level values. 
In the case of the SD  jet mass, the dynamically determined groomed jet radius $R_g$ plays the role of $R$. The term in \eq{NP} with $\df \hat\sigma_\kappa^\figeight$ is analogous to the ungroomed jet mass shift correction in the tail, but is now described by a \textit{different constant} $\Ok$ as $m^{2}_{J} =  \hat m_{J}^2 + p_TR_g \Ok$.

The term in the second line in \eq{NP} with $\df \hat\sigma_\kappa^\bndry$ is called the boundary correction. 
This effect is similar to the migration of events across $p_T$-bins due to hadronization. Near the ``boundary'' of the c-soft subjet passing/failing soft drop, i.e. when $z_g \approx \zcut r_g^\beta$, the partonic values $\hat{z}_g$ and $\hat{r}_g$ are modified due to hadronization as
\begin{align}\label{eq:zgrgShift}
	z_g = \hat z_g  + \frac{1}{r_g}\frac{\Uka}{p_TR} \, , \qquad 
	r_g  = \hat r_g - \frac{\Ukb}{p_TR} 
	\, .
\end{align}
Here, $\Uka$ characterizes the shift in the $p_T$ of the c-soft subjet analogous to jet $p_T$ shift in \eq{NPUngroomed}, and $\Ukb$ describes the change in the subjet location relative to the collinear subjet. The combination of the two gives rise to the linear structure $\Upsilon_{1\kappa}^{\bndry} = \Uka + \beta \Ukb$ as shown in \eq{NP}, and constitutes a nontrivial prediction. Finally, it is useful to factor out the parton level groomed jet mass cross section from $\df \sigma^{\figeight, \bndry}$:
\begin{align}\label{eq:C1C2Def}
	\frac{\df \hat \sigma_\kappa}{\df m_J^2}C_1^\kappa (m_J^2) \equiv 
	\frac{\df \hat \sigma^{\figeight}_\kappa}{\df m_J^2}
	\, , \quad
	\frac{\df \hat \sigma_\kappa}{\df m_J^2}
	C_2^\kappa (m_J^2) \equiv  
	\frac{\df \hat \sigma^{\bndry}_\kappa}{\df m_J^2}\,
	\,.
\end{align}
This definition is convenient as it will allow us to combine analytical calculation of the coefficients $C_{1,2}^\kappa(m_J^2)$ with parton shower jet mass cross section $\df \hat \sigma_\kappa$ as discussed below.

In \Reff{Hoang:2019ceu}, $C_{1,2}^\kappa (m_J^2)$ were computed in the coherent branching framework at LL accuracy. The first big step towards improving the accuracy of these coefficients was achieved in \Reff{Pathak:2020iue} by recasting them as moments of doubly differential cross section as in \eq{pert} and computing them at NLL$'$ accuracy in the SDOE region. 
In this work, we employ a further improved calculation at NNLL accuracy described in the companion paper in \Reff{AP}, where the matching of the doubly differential cross section in the ungroomed region is included for correct treatment of the soft drop cusp location at NNLL.
In \fig{C12Sig}, we show calculations of $\df \hat \sigma_\kappa^{\figeight,\bndry}$ for gluon jets at NNLL accuracy. With ${\cal O}(1\, {\rm GeV})$ NP constants and kinematic prefactors as shown in \eq{pert}, we see that the leading hadronization corrections can be as large as 10\% for small jet masses.

\begin{figure}[t]
	\centering
	\includegraphics[width=0.48\textwidth]{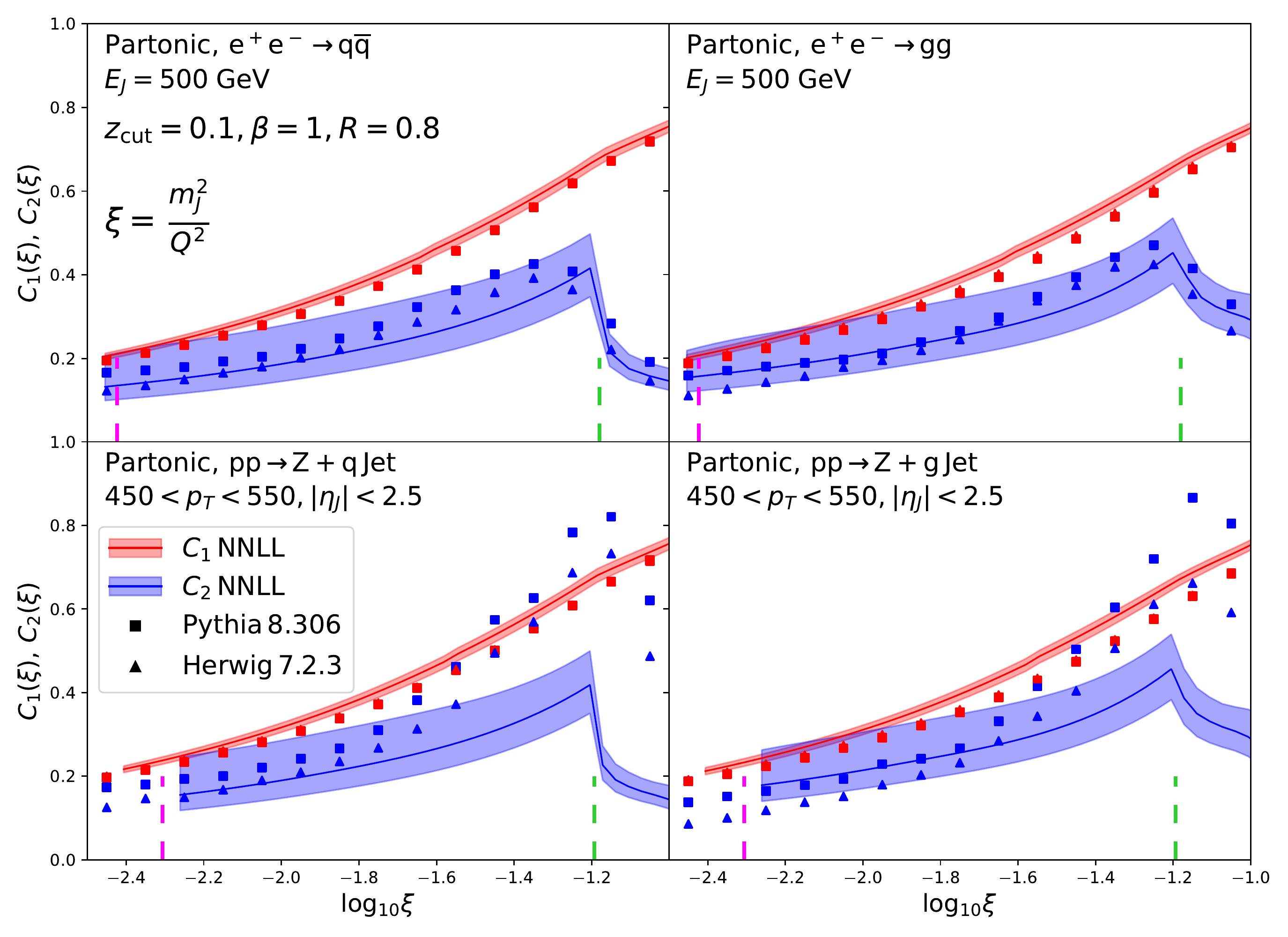}
	\caption{Weighted cross sections for hadronization corrections normalized to parton level jet mass spectrum as defined in \eq{C1C2Def} for $\zcut = 0.1$ and $\beta = 1$.}	\label{fig:C12}
\end{figure}

\emph{Calibrating hadronization models.}--- With state-of-the-art NNLL perturbative results for $C_{1,2}^\kappa(m_J^2)$, we are in position to carry out a precise calibration of hadronization models. Furthermore, by incorporating NNLL perturbative uncertainty, we are able to significantly improve upon the analysis of \Reff{Pathak:2020iue} with LL predictions lacking uncertainty estimates. We simulate $e^+e^-\ra g g$, $e^+e^-\ra q\bar q$, $pp \ra Z + q$ jet and $pp \ra Z + g$ jet processes using \Pythiaxx~\cite{Sjostrand:2007gs}, \Vinciaxx~\cite{vincia:2016} and \Herwigxx~\cite{Bahr:2008pv} parton showers with their default hadronization models. We reconstruct anti-$k_T$~\cite{Cacciari:2008gp} jets with $R=0.8$ using \texttt{Fastjet}~\cite{Cacciari:2011ma}, and analyze them using  jet analysis software \texttt{JETlib} written by two of the authors~\cite{jetlib}. For $\ee$ collisions, we sample both jets in the dijet configuration, while only using the leading jet in $pp$ collisions. As NP parameters are explictly predicted to be independent of the jet kinematics and grooming parameters, we carry out analysis using a wide range of kinematic and grooming parameter choices. In $e^+e^-$ collisions, we analyze events at center of mass energies $Q = 500,750,1000$ GeV, while in $pp$, we use jets with $p_T\in\{ [400, 600], [600,800], [800,1000]\}$ GeV and soft drop parameters $\zcut \in \{0.05, 0.1,0.15,0.2\}$ and $\beta \in \{0,0.5,1,1.5,2\}$.

We begin by explicitly defining the SDOE region where our analysis is carried out. We first define a dimensionless variable $\xi \equiv m_{J}^2/Q^2$, where
\begin{align}
    Q^{(pp)} \equiv p_T R \, , \qquad  Q^{(ee)} \equiv 2 E_J \, .
\end{align}
In terms of $\xi$, the SDOE region is then defined as
$\xi \in \big [\xi_{\rm SDOE}, \xi_0' \big] $, where
\begin{align}\label{eq:xiSDOE}
    \xi_{\rm SDOE} \equiv \xi_0 \Big(\frac{\rho \Lambda_{\rm QCD}}{Q \xi_0} \Big)^{\frac{2+\beta}{1+\beta}} \, , \quad 
    \xi_0' \equiv \frac{\xi_0}{(1 +\zeta^2)^{\frac{2+\beta}{2}}} \, .
\end{align}
Here $\xi_0$ is the location of the soft drop cusp~\cite{AP,Hannesdottir:2022rsl}: 
\begin{align}
    \xi_0^{(pp)} = \zcut \Big(\frac{R}{R_0}\Big)^{\beta} \, , \quad 
    \xi_0^{(ee)} = \zcut \bigg(\sqrt{2}\frac{\tan\frac{R}{2}}{\sin \frac{R_0}{2}}\bigg)^{\beta} \,,
\end{align}
while $\zeta$ is defined by
\begin{align}
    \zeta^{(pp)} \equiv \frac{R}{2\cosh \eta_J} \, , \qquad
    \zeta^{(ee)} \equiv \tan \frac{R}{2} \, ,
\end{align}
such that 
$\xi_0'$ in \eq{xiSDOE} is the soft-wide angle transition point of the NNLL calculation. We set $\Lambda_{\rm QCD} \ra 1$ GeV, the typical scale of transition from parton showers to hadronization. The parameter $\rho$ in \eq{xiSDOE} determines the onset of the SDOE region, and we set $\rho = 4.5$. In principle, any choice satisfying $\rho \gg 1$ is acceptable. We explore other choices of $\rho$ in the \emph{Supplemental Material}.


\begin{table}[t]
	\begin{tabular}{l | c | c | c | c }
		\hline
		\hline
		Quark Jets
		& $\Oq$(GeV)& $\Uqa$(GeV) & $\Uqb$(GeV) & $\chi^2_{\rm min}/$dof. \\ 
		\hline
		\hline
		$\ee \!\ra\! q\bar q$  & $0.55^{+0.06}_{-0.03}$  & $-0.57^{+0.16}_{-0.19}$   & $1.06^{+0.31}_{-0.35}$  & $0.77^{+0.03}_{-0.00}$      \\ \hline
		$pp\! \ra\! Z\! +\! q$ & $0.56^{+0.05}_{-0.14}$ & $-0.73^{+0.29}_{-0.28}$ & $0.89^{+0.27}_{-0.25}$ & $0.65^{+0.01}_{-0.02}$\\
		\hline
	\end{tabular}	
\vspace{5pt}

\begin{tabular}{l | c | c | c | c }
		\hline	 \hline 
		Gluon Jets
		& $\Og$(GeV)& $\Uga$(GeV) & $\Ugb$(GeV) & $\chi^2_{\rm min}/$dof. \\ 
		\hline\hline
		$e^+e^-\! \ra\! gg$ & $1.92^{+0.16}_{-0.32}$ & $-0.48^{+0.23}_{-0.22}$ & $0.87^{+0.25}_{-0.25}$ & $3.13^{+0.05}_{-0.20}$\\ 
		\hline
		$pp\! \ra\!Z\! +\! g$ & $0.93^{+0.01}_{-0.12}$ & $-0.24^{+0.11}_{-0.01}$ & $0.89^{+0.20}_{-0.23}$ & $1.34^{+0.05}_{-0.10}$ \\
		\hline
	\end{tabular}
	\caption{\label{tab:PythiaFit} Fit results for NP constants in \Pythiaxx for quark and gluon jets in $\ee$ and $pp$ collisions.}
\end{table}


In \fig{C12} we show a comparison of the NNLL computation of  $C_{1,2}^\kappa$ with partonic \Pythia and \Herwig. The  parton level results for from \Vincia are found to be almost identical to \Pythia. We find a good agreement of the NNLL $C_1^\kappa$ with MC for all four processes. The unusually small errors for $C_1^\kappa$ result from cancellation between correlated uncertainties in the two factors in \eq{C1C2Def}. 
For $pp$, the agreement for the boundary term is poor for jet masses close to the cusp due to the initial-state radiation (ISR) contribution. However, as seen in \fig{C12Sig}, the NP corrections in the cusp-region are relatively suppressed, and NP corrections from ISR are also expected to be smaller as they involve subleading $r_g^2$ moment of the boundary cross section~\cite{AP}. Consequently, these effects do not significantly impact the analysis below.

Finally, we perform a least-squares fit for the NP parameters by defining our $\chi^2$ statistic as
\begin{align}
    \chi^2  \equiv  \sum_{i} \frac{\big[(\vec \sigma^{\rm MC}_{\kappa,\rm had})_i - (\vec  \sigma_{\kappa, \rm part + NP}(\Ok,\ldots)\big)_i \big]^2}{(\Delta\vec  \sigma)^2_i}.
\end{align}
Here, $\vec \sigma_X$ is a vector of cross section values for $n_{\rm bins} = 10$ bins in the fit range and all permutations of $p_T$ (or $E_J$), $\zcut$, and $\beta$ values considered above. We denote the hadron level MC groomed jet mass cross section as $\vec \sigma^{\rm MC}_{\kappa,\rm had}$, and define $\vec  \sigma_{\kappa,\rm part + NP}$ by including the NP constants $\Ok, \Uka, \Ukb$ and NNLL computation of $C_{1,2}^\kappa$ in \eq{C1C2Def} to the parton level MC spectrum $\df\hat{\sigma}^{\rm MC}_{\kappa}$
following \eq{NP}. 
The uncertainty in the denominator is defined as
\begin{align}\label{eq:errChi2}
    (\Delta\vec  \sigma)_i^2 \equiv \big(0.05 (\vec  \sigma_{\rm part \times C_1})_i\big)^2 + \big(0.25 (\vec  \sigma_{\rm part \times C_2})_i\big)^2\, , 
\end{align}
where, guided by the size of perturbative uncertainties in \fig{C12Sig}, we have assigned 5\% and 25\% uncertainty respectively to the weighted cross sections for shift and boundary corrections respectively. The NP constants $\Ok, \Uka, \Ukb$ are then varied to minimize this $\chi^2$ statistic. An example of the fit for mass distribution is shown in \fig{PythiaFit}.

\begin{figure}[t]
	\centering
	\includegraphics[width=0.35\textwidth]{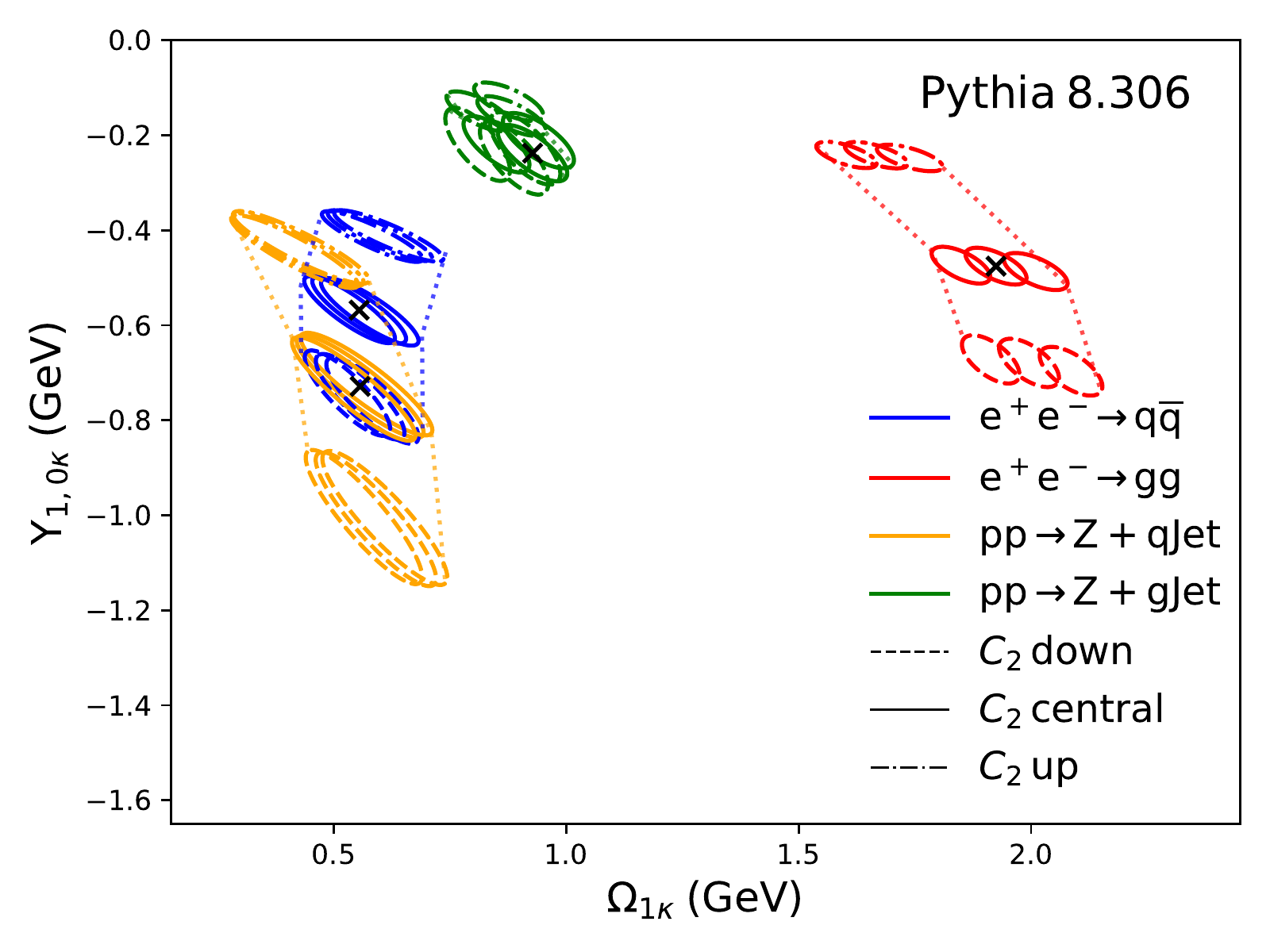}
	\includegraphics[width=0.48\textwidth]{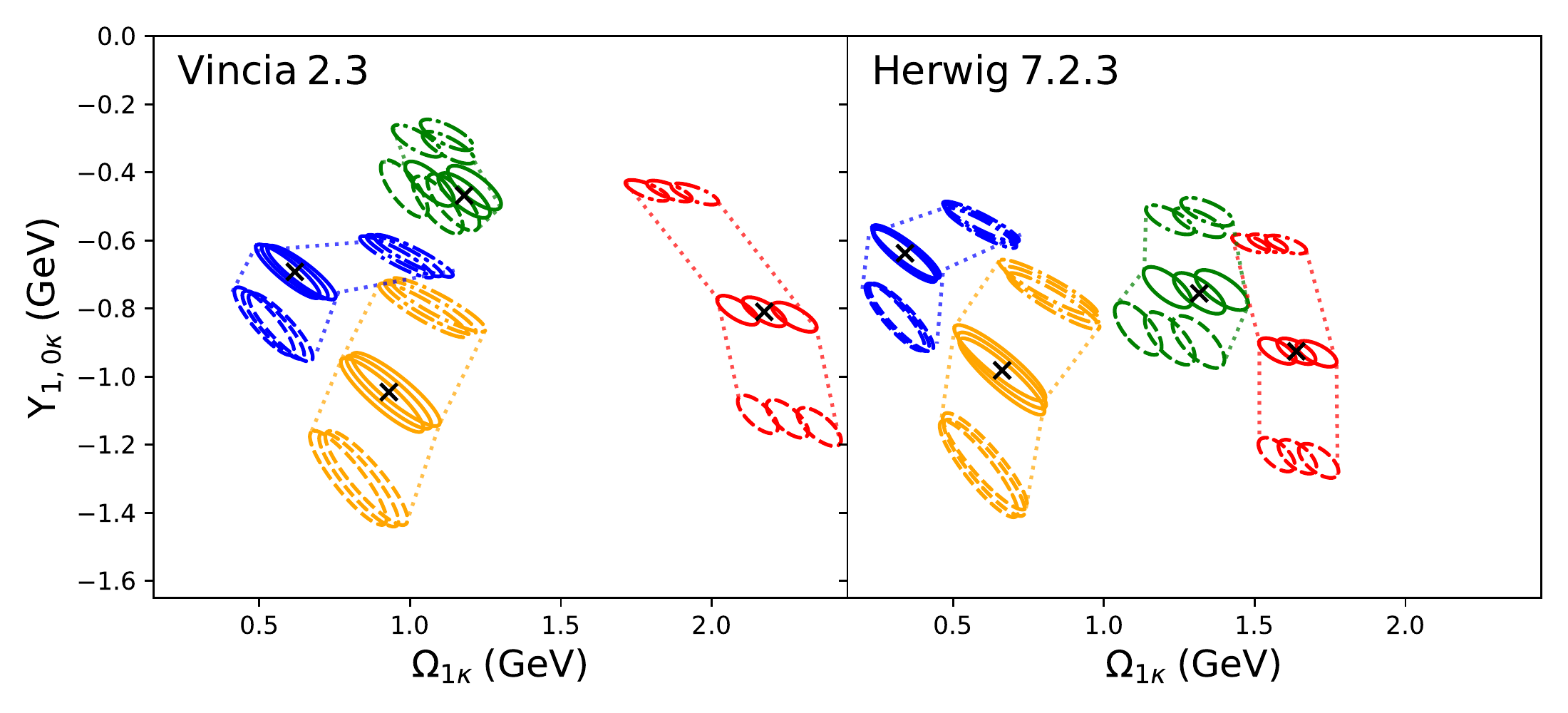}
	\caption{Testing jet flavor universality of soft drop NP parameters in \Pythiaxx (top), \Vinciaxx (bottom, left) and \Herwigxx (bottom, right).}
	\label{fig:contours}
\end{figure}

In \tab{PythiaFit}, we present the fit results for the NP constants with scale variations of $C_{1,2}^\kappa$ for \Pythia. As anticipated, the parameters are $\lesssim 1$ GeV. We also find similar parameter values for quark jets within the two quark processes within uncertainties. Even when NP parameters for quark jets are simultaneously fit for in $e^+e^-$ and $pp$ process, we find an excellent $\chi^2$ value of $0.840/$dof. This is expected, as soft drop isolates the jet from surrounding radiation. To further investigate this, we show correlations between $\Ok$ and $\Uka$ for the four processes in \fig{contours} where each ellipse represents a 1$\sigma$ deviation. To account for perturbative uncertainties, we repeat the fit by varying $C_{1,2}^\kappa$ up and down within the uncertainty band shown in \fig{C12}. We observe an excellent agreement within uncertainties between the NP parameters for quark jets in $pp$ and $\ee$ collisions in \Pythia simulations, and a moderate agreement for \Vincia and \Herwig. In contrast, while \Herwig exhibits similar levels of agreement for gluon jets and quark jets at both colliders, \Pythia and \Vincia show significant disagreement. This shows that contrary to the expectation for groomed jets, hadronization modeling of gluon jets in isolation in $\ee$ collisions in \Pythia and \Vincia differs significantly from jets in hadron colliders. Additionally, the differing results between \Pythia and \Vincia point to the interplay of parton showers with hadronization models. In the \emph{Supplemental Material} we show correlations in the other two combinations of NP parameters which show similar behavior as well as numerical fit results for \Herwig and \Vincia.

\begin{figure}[]
\centering
\includegraphics[width=0.48\textwidth]{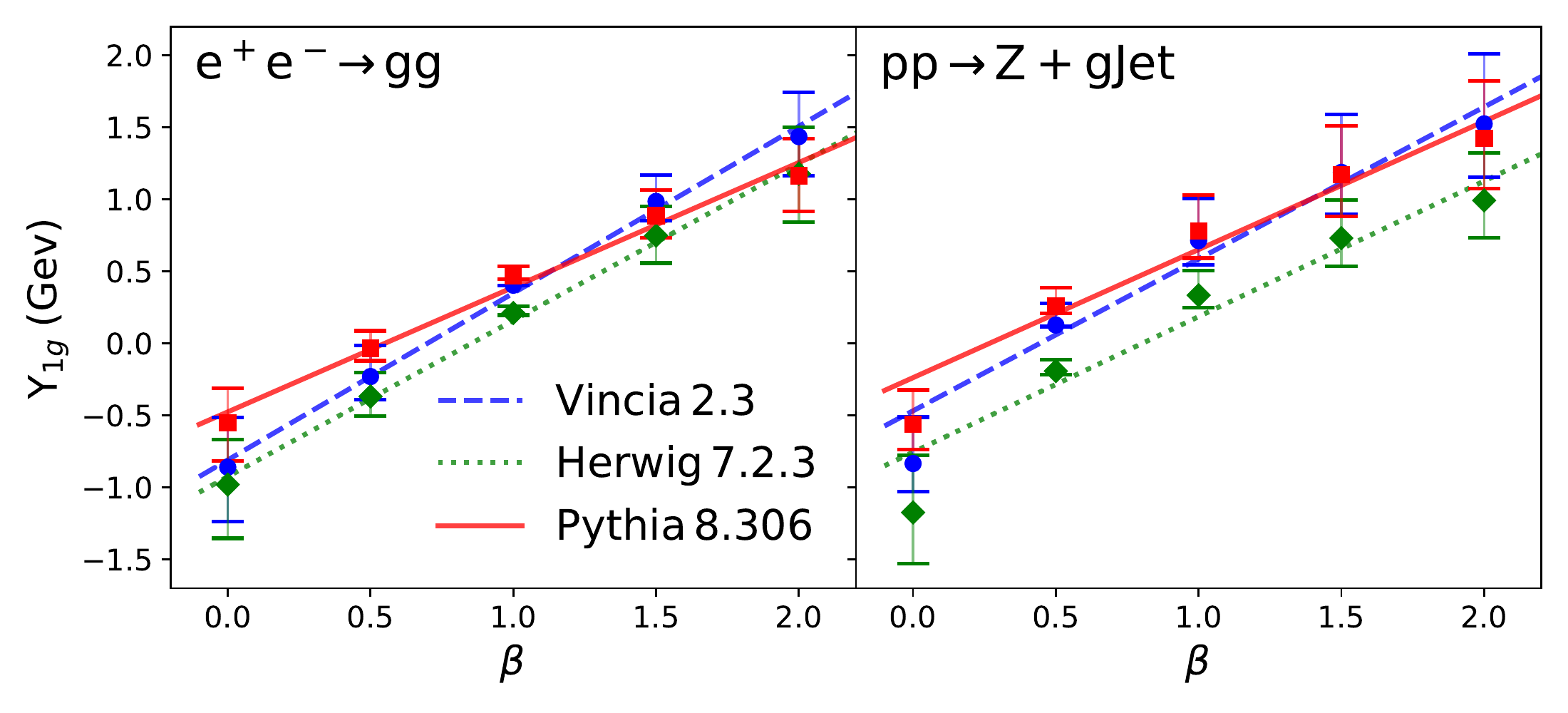}
\includegraphics[width=0.48\textwidth]{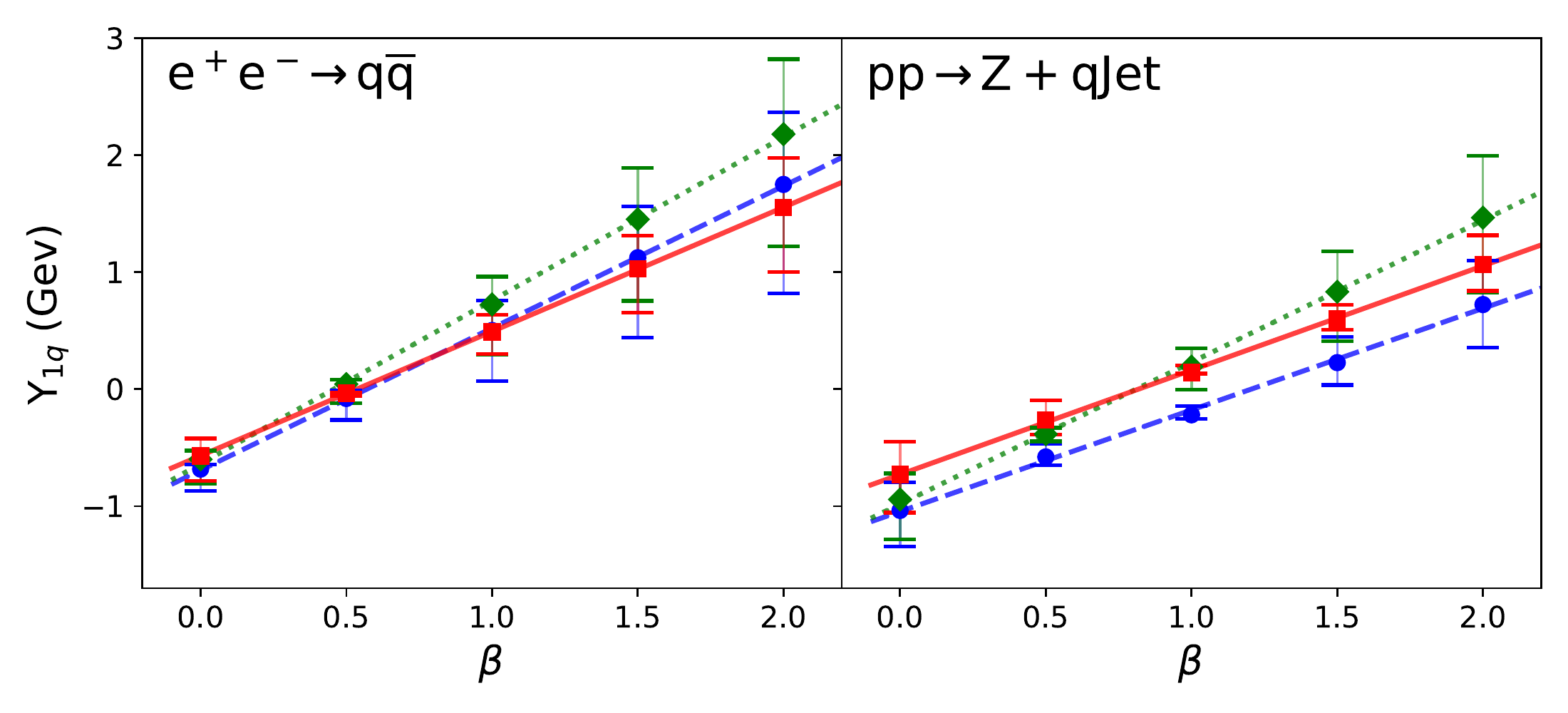}
\caption{Test for linear $\beta$ dependence of boundary corrections ($\Upsilon_{1\kappa}^{\bndry} = \Uka + \beta \Ukb$) in gluon (top) and quark (bottom) jets for $\ee$ (left) and $pp$ (right) collisions}
\label{fig:split_ups}
\end{figure}

\begin{figure}[t]
	\centering
	\includegraphics[width=0.48\textwidth]{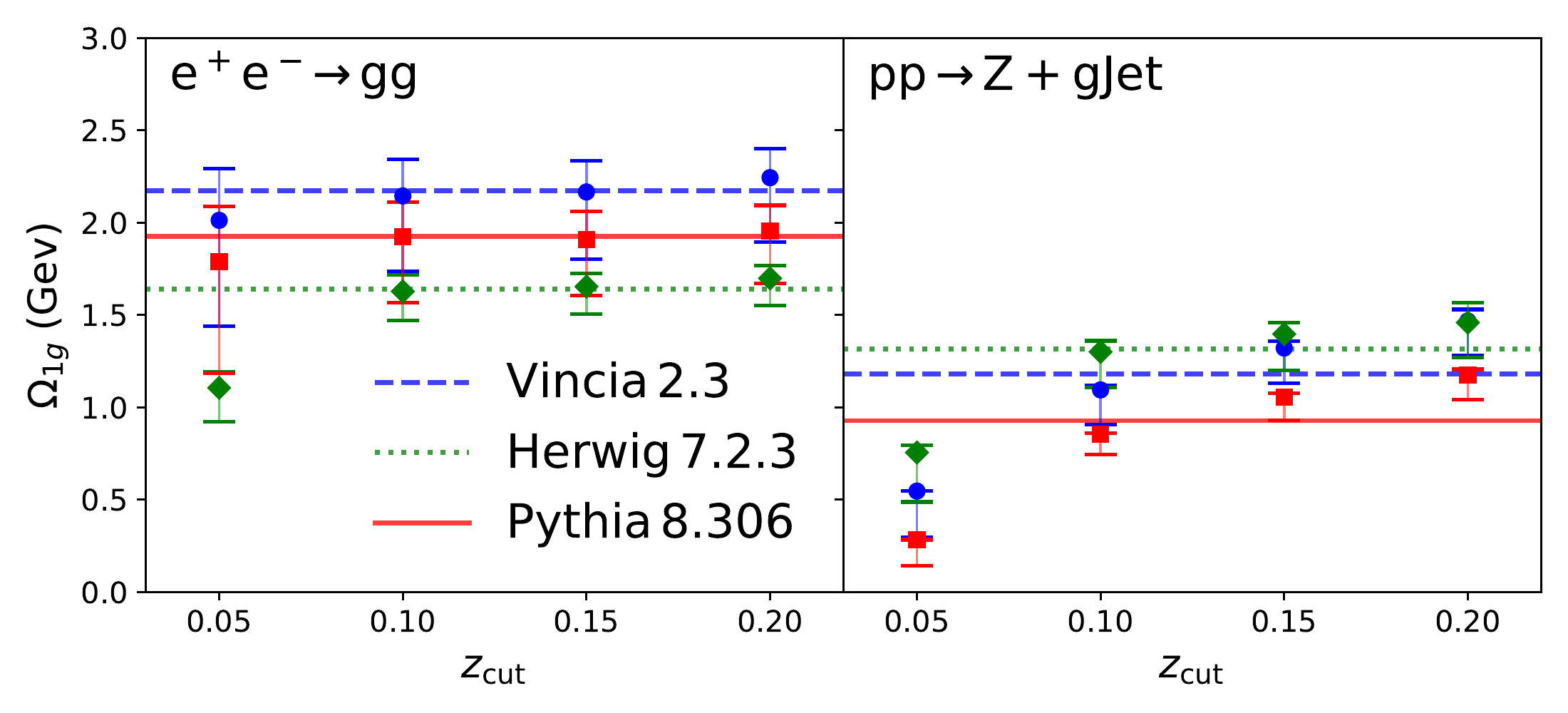}
	\includegraphics[width=0.48\textwidth]{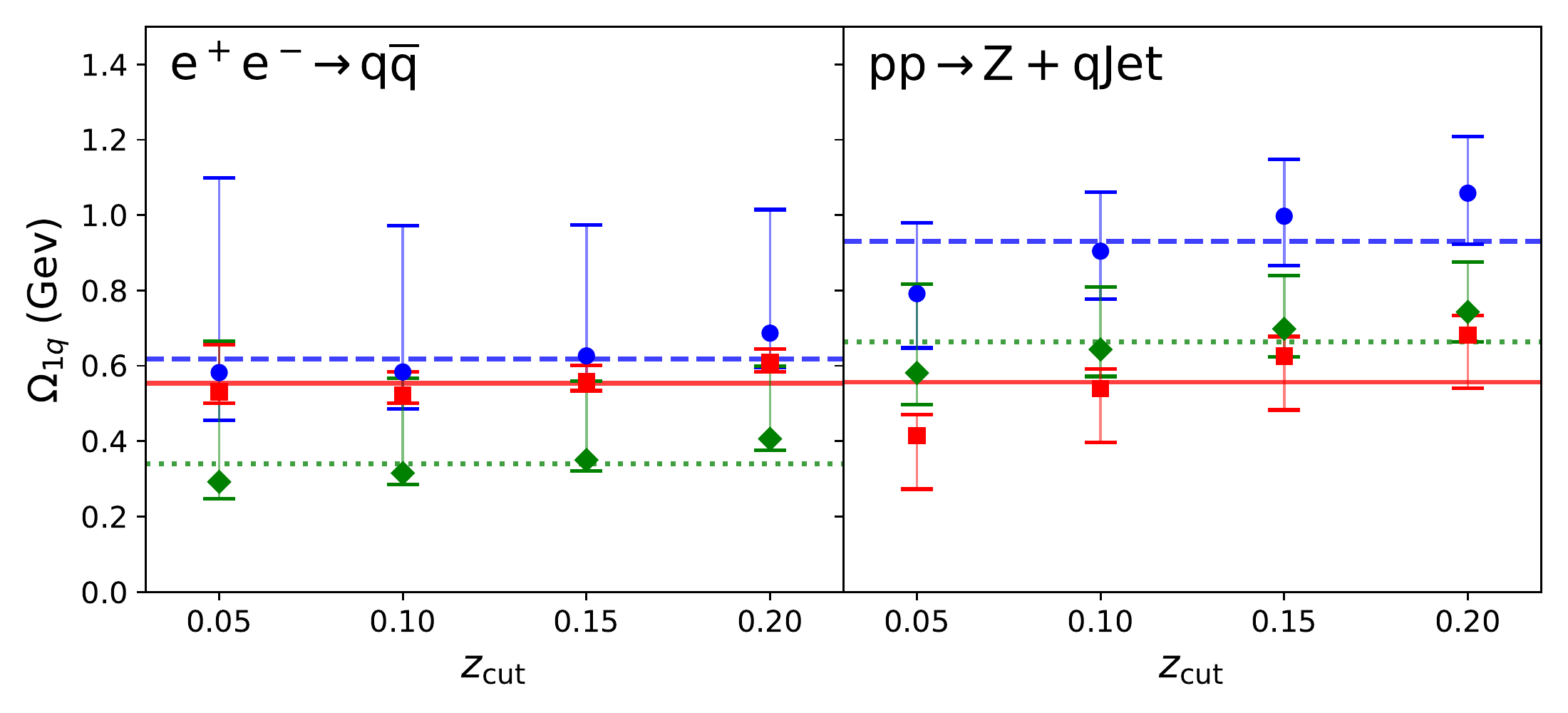}
	\caption{Testing $\zcut$-independence of $\Ok$ for gluon (top) and quark jets (bottom) in $\ee$ (left) and $pp$ collisions.}
	\label{fig:o1_zcut}
\end{figure}

Next, we test the grooming parameters independence of these NP constants. We follow the same procedure as \Reff{Hoang:2019ceu} and test this behavior by comparing the fit results for individual $\zcut$ and $\beta$ values with the global fit. In \fig{split_ups}, we demonstrate the linear $\beta$-dependence of the boundary correction by fitting for a single parameter $\Upsilon_{1\kappa}^\bndry(\beta)$ for each value of $\beta$. Because of degeneracy in the NP parameters, we fix $\Ok$ to its global-fit value in this case. The error bars take into account perturbative uncertainty in $C_{1,2}^\kappa$ by re-fitting with minimum and maximum variations. 
We find that all the three simulations perform well in each of the four cases. 
In \fig{o1_zcut}, we repeat the same procedure to test $\zcut$-independence of NP parameters. We find here that the three event generators pass the test for both quark and gluon jets in $\ee$ collisions, but exhibit a linear trend in $\zcut$ for both flavors in $pp$ collisions. 
The larger $\chi^2$ values for gluon jets, as seen in \tab{PythiaFit} for \Pythia (also true for \Herwig and \Vincia) suggest that modeling of hadronization in gluon jets is less consistent with our field theory predictions.
Finally, our analysis of the $\ee \ra q \bar q$ process using NNLL predictions of $C_{1,2}^\kappa$ demonstrates significant improvement in the universality behavior of $\zcut$ and $\beta$, compared to \Reff{Hoang:2019ceu} where LL predictions were used.\footnote{Note that our numerical results for $\ee \ra q \bar q$ also differ from those in \Reff{Hoang:2019ceu} due to different prescription for error in \eq{errChi2} and newer versions of MC.}
In conclusion, while 
our universality tests of the NP parameters generally display expected behaviors in all the cases considered, they also reveal some tension with the hadronization models, pointing to interesting avenues for further improvement\footnote{For example, the analysis at LL in Reference~\cite{Hoang:2019ceu} already revealed problems in the \Herwig \texttt{8.2} hadronization model, which resulted in its improvement in version \texttt{8.3}.} and motivate the use of real-world collider data for further analyses.

\emph{Conclusions.}---
In this \emph{Letter}, we have presented a systematic framework for analyzing nonperturbative corrections in soft drop jet mass by bringing together earlier work on nonperturbative factorization and high precision calculations of multi-differential soft drop cross sections.  Our analysis with hadronization models successfully demonstrates that nonperturbative parameters exhibit the universal behaviors predicted by field theory.
Our analysis is also directly applicable to precision phenomenology involving soft drop jet mass. For example, in \Reff{Hannesdottir:2022rsl} our results are used  to assess the impact of the NP corrections on the sensitivity and ultimate precision achievable on $\as$ at the LHC using SD jet mass. Findings in \Reff{Hannesdottir:2022rsl} indicate that the hadronization effects in the $\beta = 1$ case, for instance, are 3\% (8\%) for quark (gluon) jets when nonperturbative parameters in \eq{NP} are left unconstrained, which are of the same size as the NNLL perturbative uncertainty. We anticipate that with high precision calculations for the soft drop jet mass and the boundary correction ($C_2^\kappa$ in \fig{C12}), it will be possible to significantly constrain some or all of the NP constants, and hence improve the ultimate precision achievable on $\as$-determination at the LHC.
In summary, our work thus provides crucial understanding of hadronization corrections necessary for precision measurements with soft drop jet mass, a benchmark tool for improving hadronization modeling in MC event generators, and motivation for analyses with real world collider data.

\emph{Acknowledgements.}--- 
We would like to thank Mrinal Dasgupta, Michael Seymour for helpful discussions. 
We are grateful to Simon Pl\"atzer for many discussions and support with analysis with \Herwig. We thank Holmfridur Hannesdottir, Johannes Michel and Iain Stewart for numerous discussions and feedback on the manuscript. We provide a numerical implementation of the NNLL calculation in \texttt{C++} building on core classes of \texttt{SCETlib}~\cite{scetlib} which will be made available as a part of the \texttt{scetlib::sd} module~\cite{scetlibSD}. We thank Johannes Michel for support with above-mentioned implementation in \texttt{SCETlib}. KL was supported by the LDRD program of LBNL and the U.S. DOE under contract number DE-SC0011090. AP acknowledges support from DESY (Hamburg, Germany), a member of the Helmholtz Association HGF. AP was a member of the Lancaster-Manchester-Sheffield Consortium for Fundamental Physics, which is supported by the UK Science and Technology Facilities Council (STFC) under grant number ST/T001038/1. AF also gratefully acknowledges support from the above-mentioned grant.

\bibliography{sd}
\end{document}


\section*{Supplemental Material}
\label{sec:supplemental}

\begin{figure}[h]
	\centering
	\includegraphics[width=0.48\textwidth]{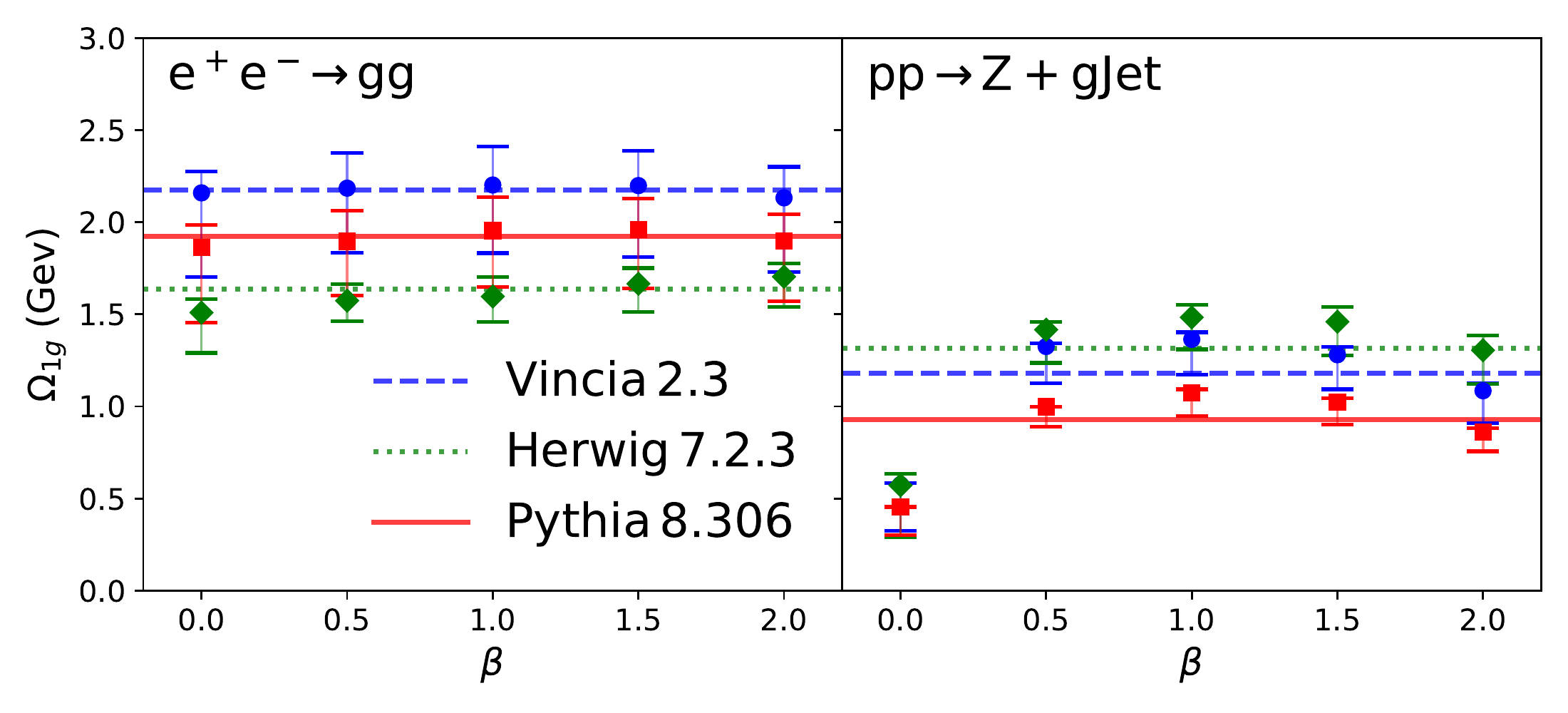}
	\includegraphics[width=0.48\textwidth]{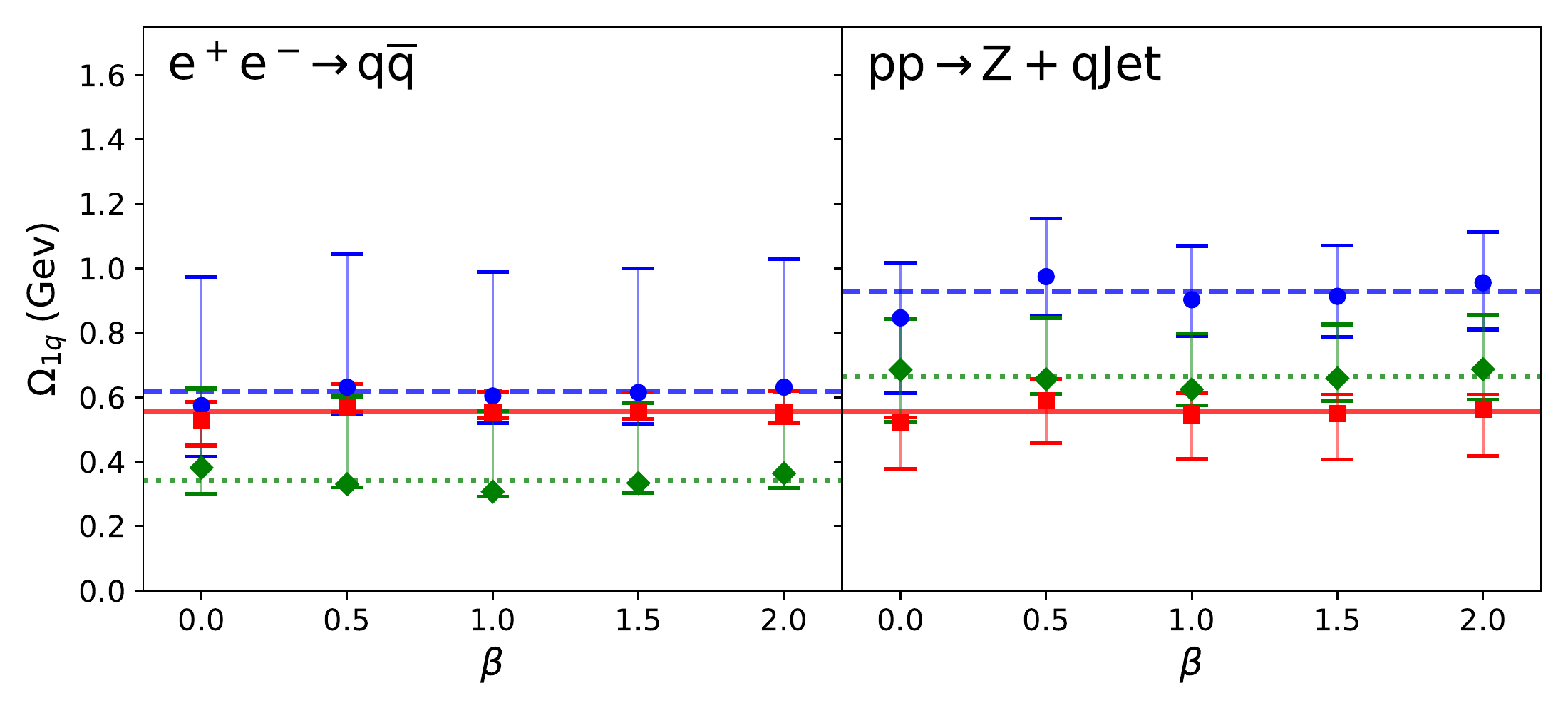}
	
	\caption{Testing $\beta$-independence of $\Ok$ for quark and gluon jets while fixing $\Uka$ and $\Ukb$ paramaters to their global-fit values.}
	\label{fig:o1_beta}
\end{figure}

\begin{table}[!htb]
		\centering
		\begin{tabular}{l | c | c | c | c }
			\hline
			\hline
			Quark Jets
			& $\Oq$(GeV)& $\Uqa$(GeV) & $\Uqb$(GeV) & $\chi^2_{\rm min}/$dof. \\ 
			\hline
			\hline
			$\ee \!\ra\! q\bar q$  & $0.62^{+0.39}_{-0.10}$ & $-0.69^{+0.05}_{-0.16}$ & $1.21^{+0.39}_{-0.48}$ & $0.94^{+0.06}_{-0.01}$   \\ \hline
			$pp\! \ra\! Z\! +\! q$ &$0.93^{+0.16}_{-0.14}$ & $-1.05^{+0.26}_{-0.26}$ & $0.87^{+0.32}_{-0.30}$ & $0.82^{+0.02}_{-0.02}$	\\
			\hline
		\end{tabular}	
		\vspace{5pt}

		\begin{tabular}{l | c | c | c | c }
			\hline	 \hline 
			Gluon Jets
			& $\Og$(GeV)& $\Uga$(GeV) & $\Ugb$(GeV) & $\chi^2_{\rm min}/$dof. \\ 
			\hline\hline
			$e^+e^-\! \ra\! gg$ & $2.17^{+0.18}_{-0.39}$ & $-0.81^{+0.36}_{-0.34}$ & $1.16^{+0.34}_{-0.32}$ & $3.56^{+0.07}_{-0.20}$\\ 
			\hline
			$pp\! \ra\!Z\! +\! g$ & $1.18^{+0.03}_{-0.20}$ & $-0.47^{+0.18}_{-0.03}$ & $1.06^{+0.24}_{-0.28}$ & $1.75^{+0.08}_{-0.10}$  \\ \hline
		\end{tabular}
		\caption{\label{tab:VinciaFit} Fit results for NP constants in \Vinciaxx.}
		\centering
		\begin{tabular}{l | c | c | c | c }
			\hline
			\hline
			Quark Jets
			& $\Oq$(GeV)& $\Uqa$(GeV) & $\Uqb$(GeV) & $\chi^2_{\rm min}/$dof. \\ 
			\hline
			\hline
			$\ee \!\ra\! q\bar q$  &  $0.34^{+0.25}_{-0.03}$ & $-0.64^{+0.09}_{-0.19}$ & $1.40^{+0.42}_{-0.51}$ & $0.63^{+0.10}_{-0.02}$   \\ \hline
			$pp\! \ra\! Z\! +\! q$ & $0.66^{+0.17}_{-0.08}$ & $-0.98^{+0.24}_{-0.29}$ & $1.21^{+0.41}_{-0.42}$ & $0.27^{+0.03}_{-0.01}$	\\
			\hline
		\end{tabular}	
		\vspace{5pt}
		
		\begin{tabular}{l | c | c | c | c }
			\hline	 \hline 
			Gluon Jets
			& $\Og$(GeV)& $\Uga$(GeV) & $\Ugb$(GeV) & $\chi^2_{\rm min}/$dof. \\ 
			\hline\hline
			$e^+e^-\! \ra\! gg$ & $1.64^{+0.07}_{-0.15}$ & $-0.93^{+0.32}_{-0.32}$ & $1.09^{+0.34}_{-0.33}$ & $1.70^{+0.02}_{-0.02}$\\ 
			\hline
			$pp\! \ra\!Z\! +\! g$ & $1.32^{+0.07}_{-0.20}$ & $-0.76^{+0.24}_{-0.14}$ & $0.94^{+0.23}_{-0.26}$ & $0.71^{+0.07}_{-0.03}$ \\
			\hline
		\end{tabular}
		\caption{\label{tab:HerwigFit} Fit results for NP constants in \Herwigxx.}
\end{table}
In this \emph{Supplemental material}, we present additional results for the calibration exercise of the MC hadronization model. Tables \ref{tab:VinciaFit} and \ref{tab:HerwigFit} present the results of fitting to \Vinciaxx and \Herwigxx, respectively. Similar to fits to \Pythiaxx in Tab.~I discussed in the main text, the $\chi^2$ values in \tab{VinciaFit} and \tab{HerwigFit} show that the fits for quark jets are much better constrained than gluon jets for both \Vincia and \Herwig, with those of \Herwig being more consistent with our predictions.

\begin{figure}[t]
	\centering
	\includegraphics[width=0.48\textwidth]{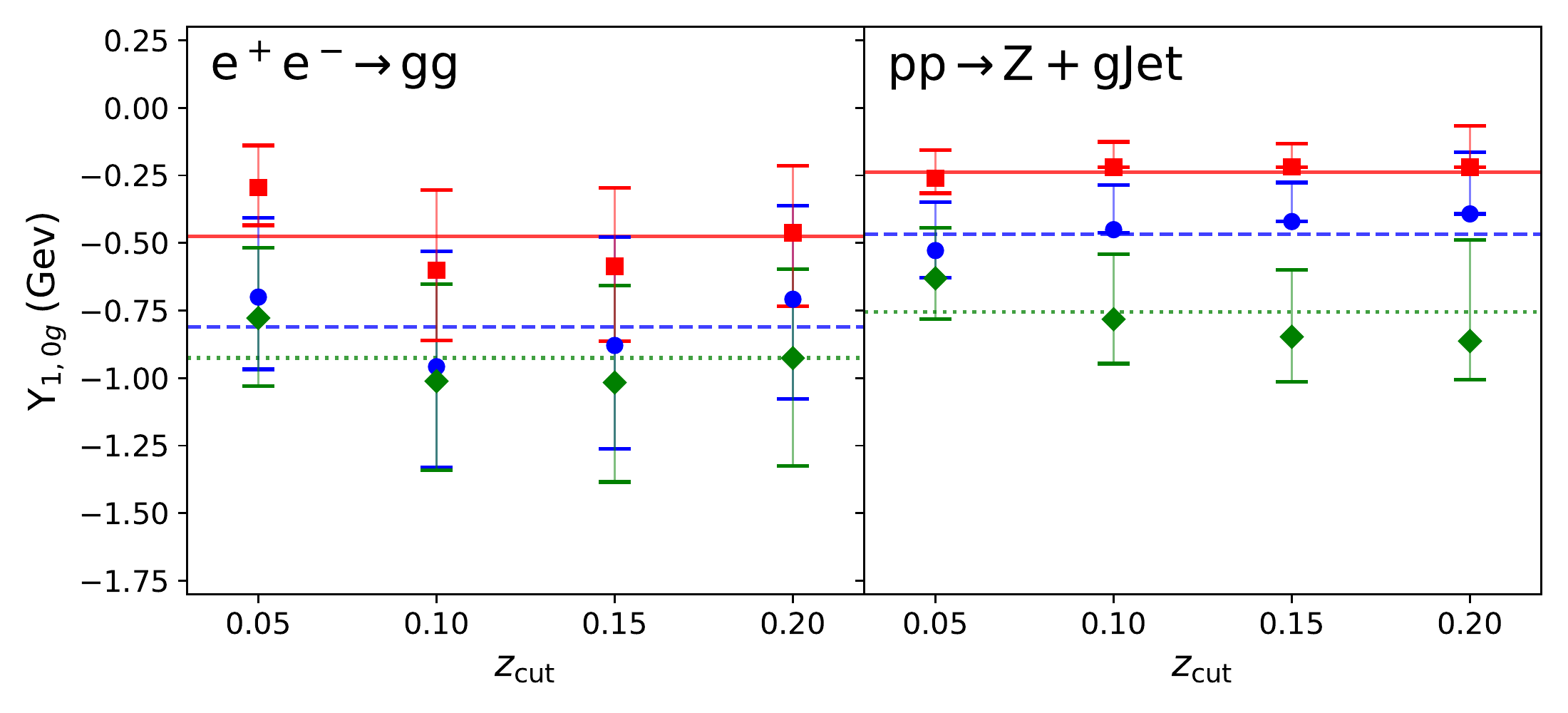}
	\includegraphics[width=0.48\textwidth]{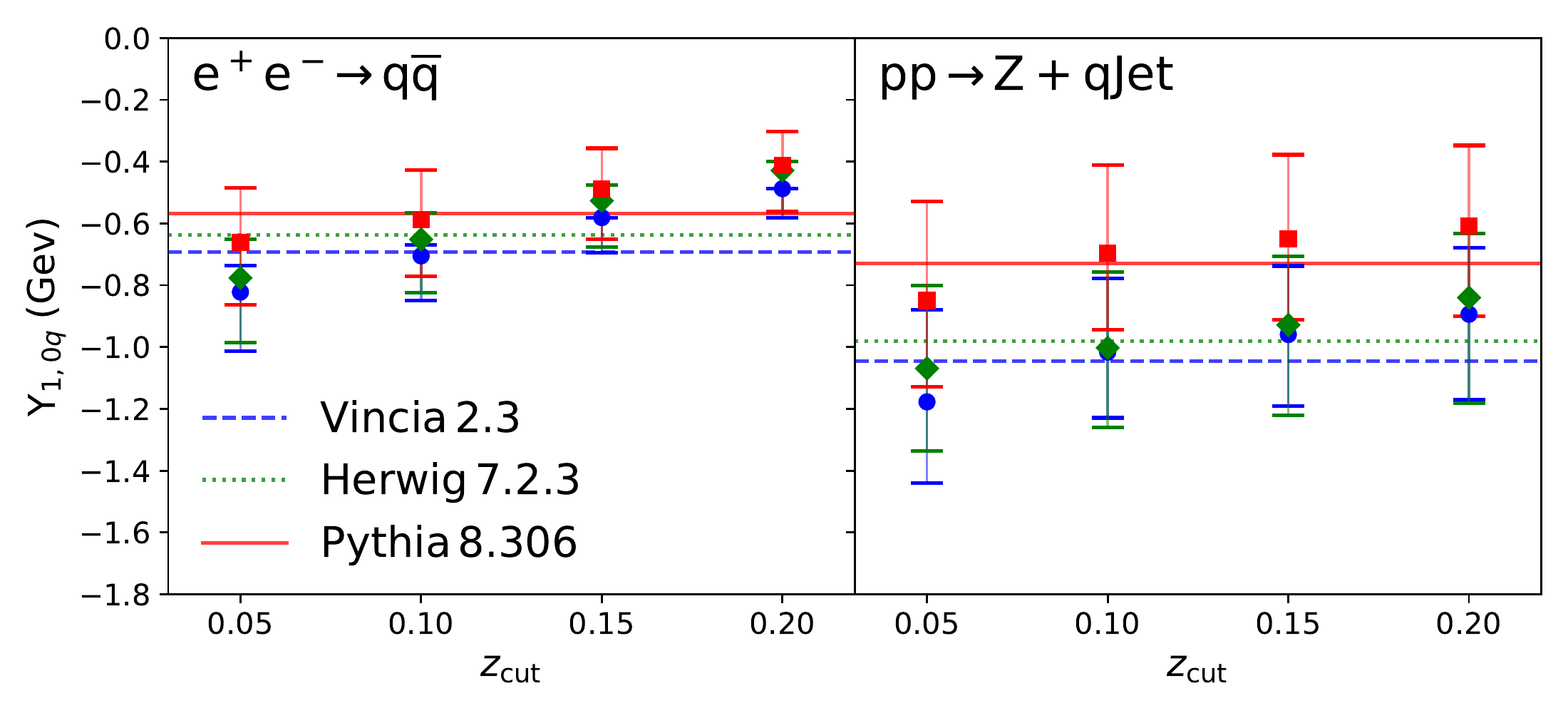}
	
	\caption{Testing $\zcut$-independence of $\Uka$ for quark and gluon jets while fixing $\Ukb$ and $\Ok$ paramaters to their global-fit values.}
	\label{fig:uka_zcut}
\end{figure}

\begin{figure}[t]
	\centering
	\includegraphics[width=0.48\textwidth]{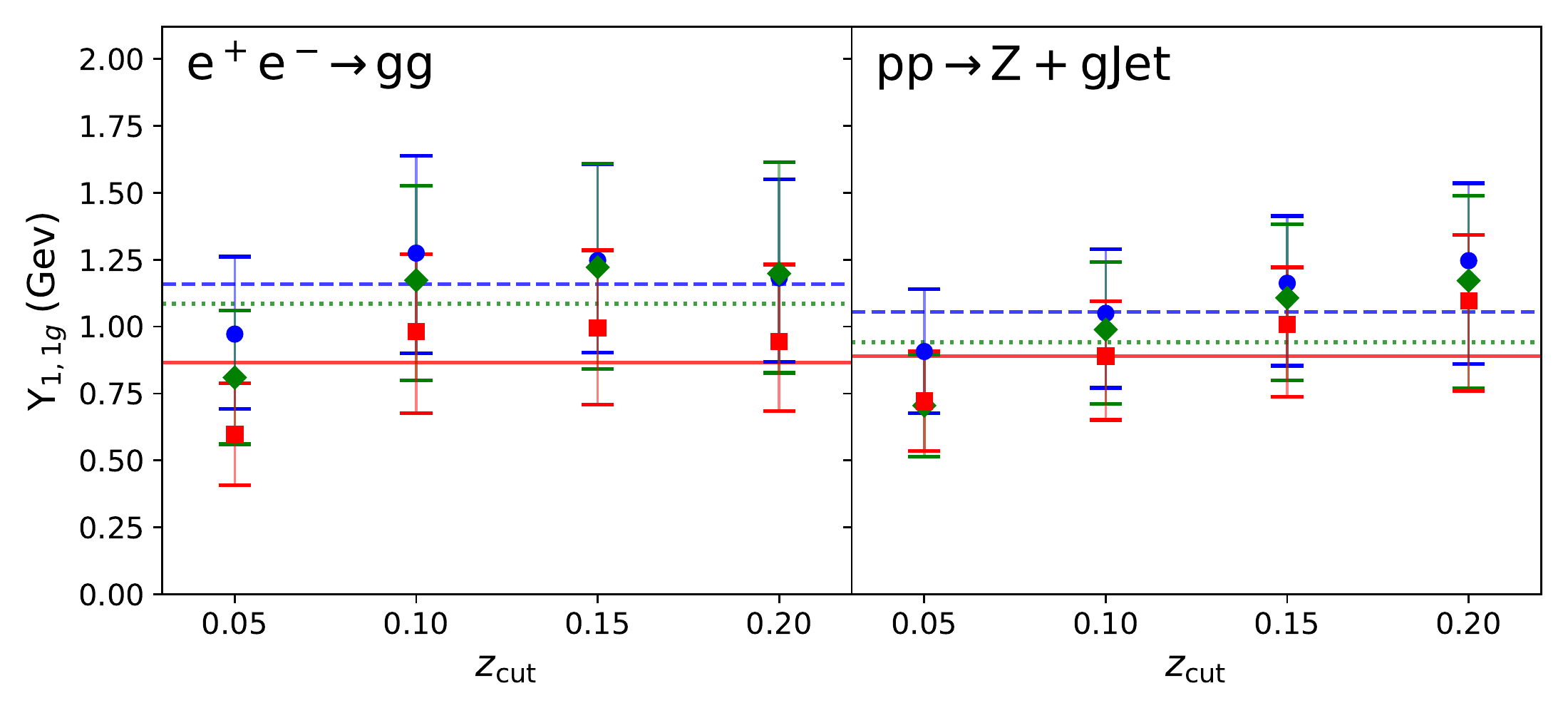}
	\includegraphics[width=0.48\textwidth]{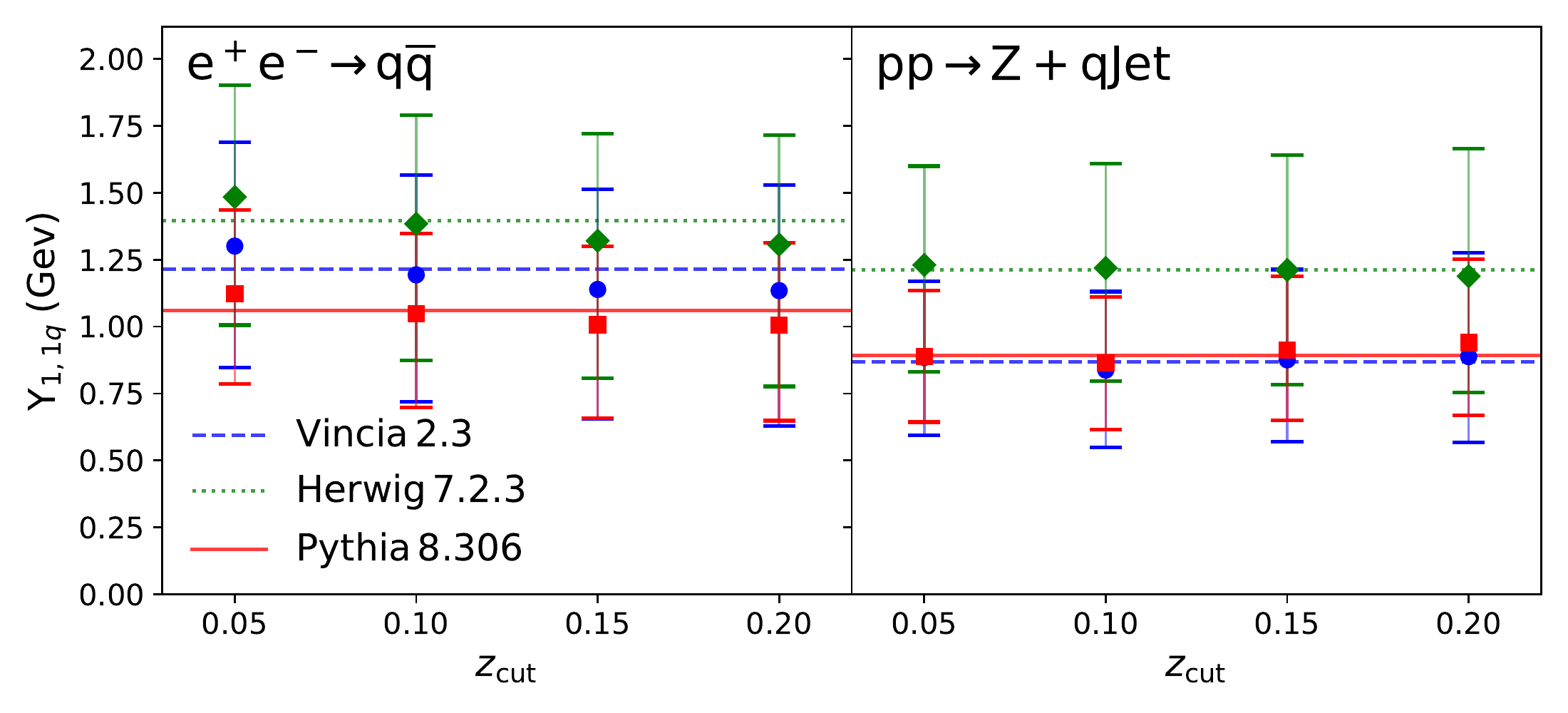}
	
	\caption{Testing $\zcut$-independence of $\Ukb$ for quark and gluon jets while fixing $\Uka$ and $\Ok$ paramaters to their global-fit values.}
	\label{fig:ukb_zcut}
\end{figure}
In Figs.~\ref{fig:o1_beta}, \ref{fig:uka_zcut} and \ref{fig:ukb_zcut} we show tests for the independence of the grooming parameters $\Ok$, $\Uka$, and $\Ukb$. The horizontal lines in these figures represent the global-fit values, while the markers with error bars represent fits for individual $\zcut$ or $\beta$ parameters. 
 We observe that for each process the results of fits to individual $\zcut$ or $\beta$ values are consistent with the global-fit value within uncertainty, with the exception of  $\beta = 0$ result of $\Og$ for gluon jets in \fig{o1_beta} where we notice a significant deviation.

\begin{figure}[t]
	\centering
	\includegraphics[width=0.45\textwidth]{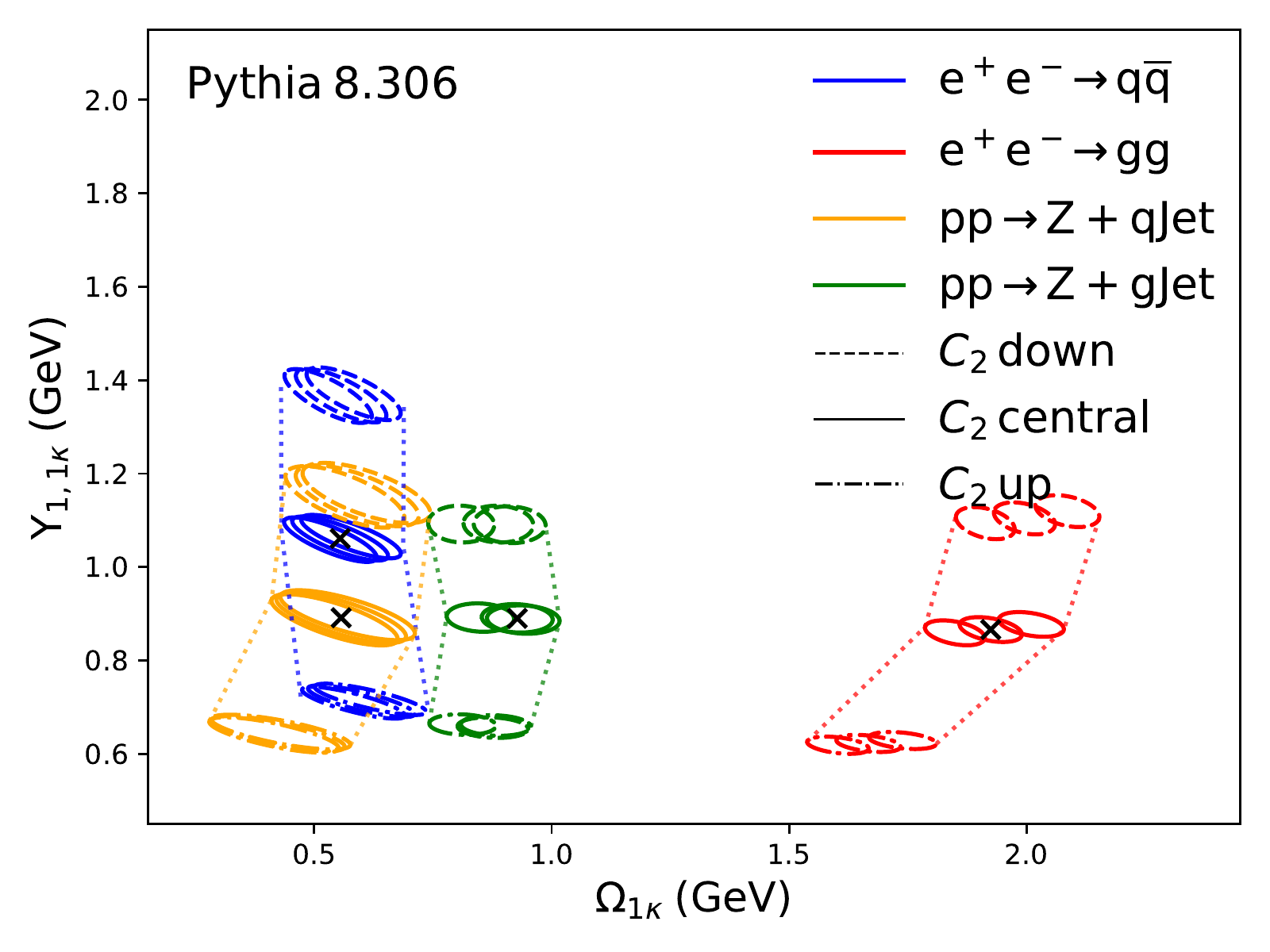}
	\includegraphics[width=0.45\textwidth]{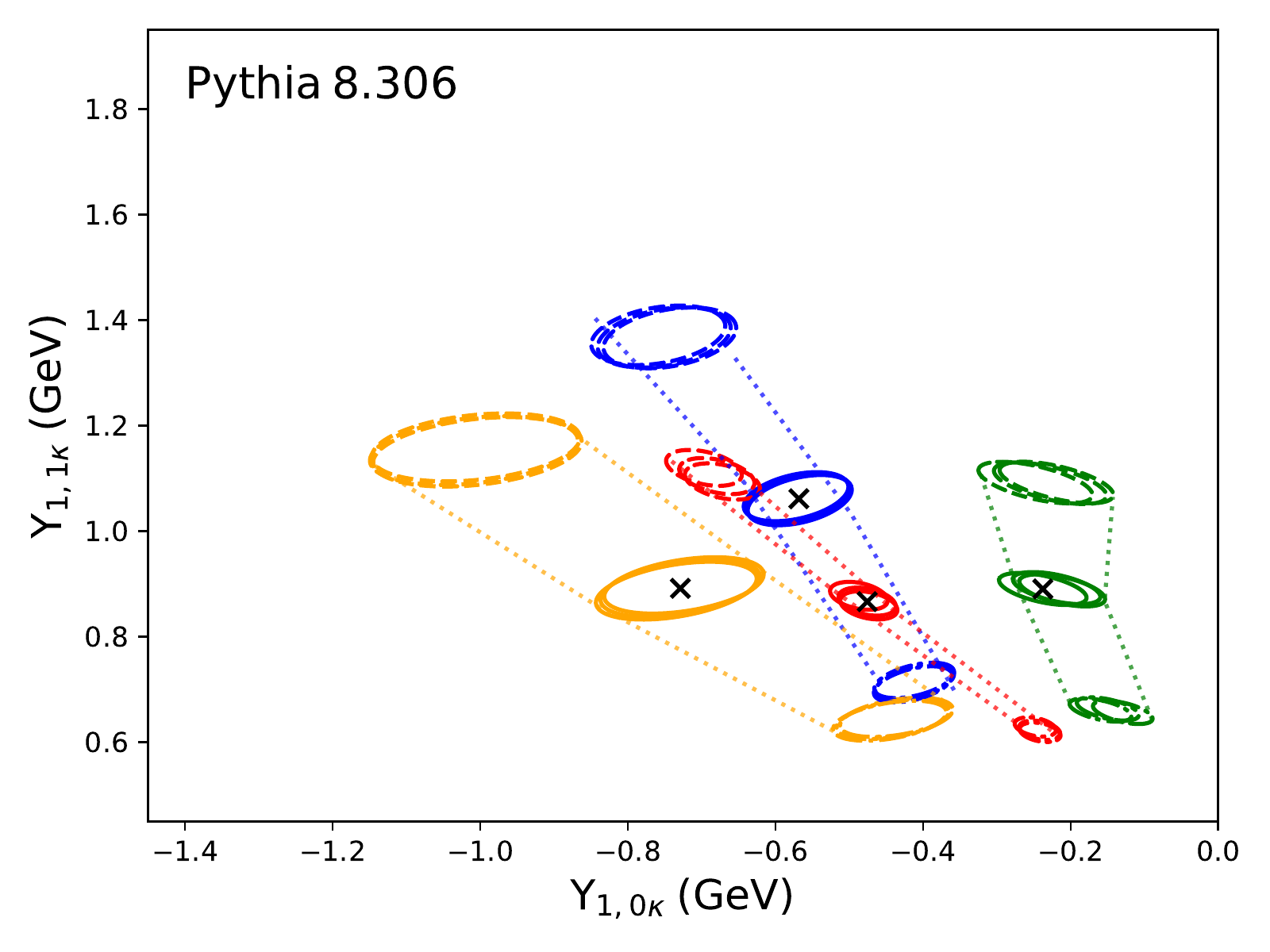}
	\includegraphics[width=0.45\textwidth]{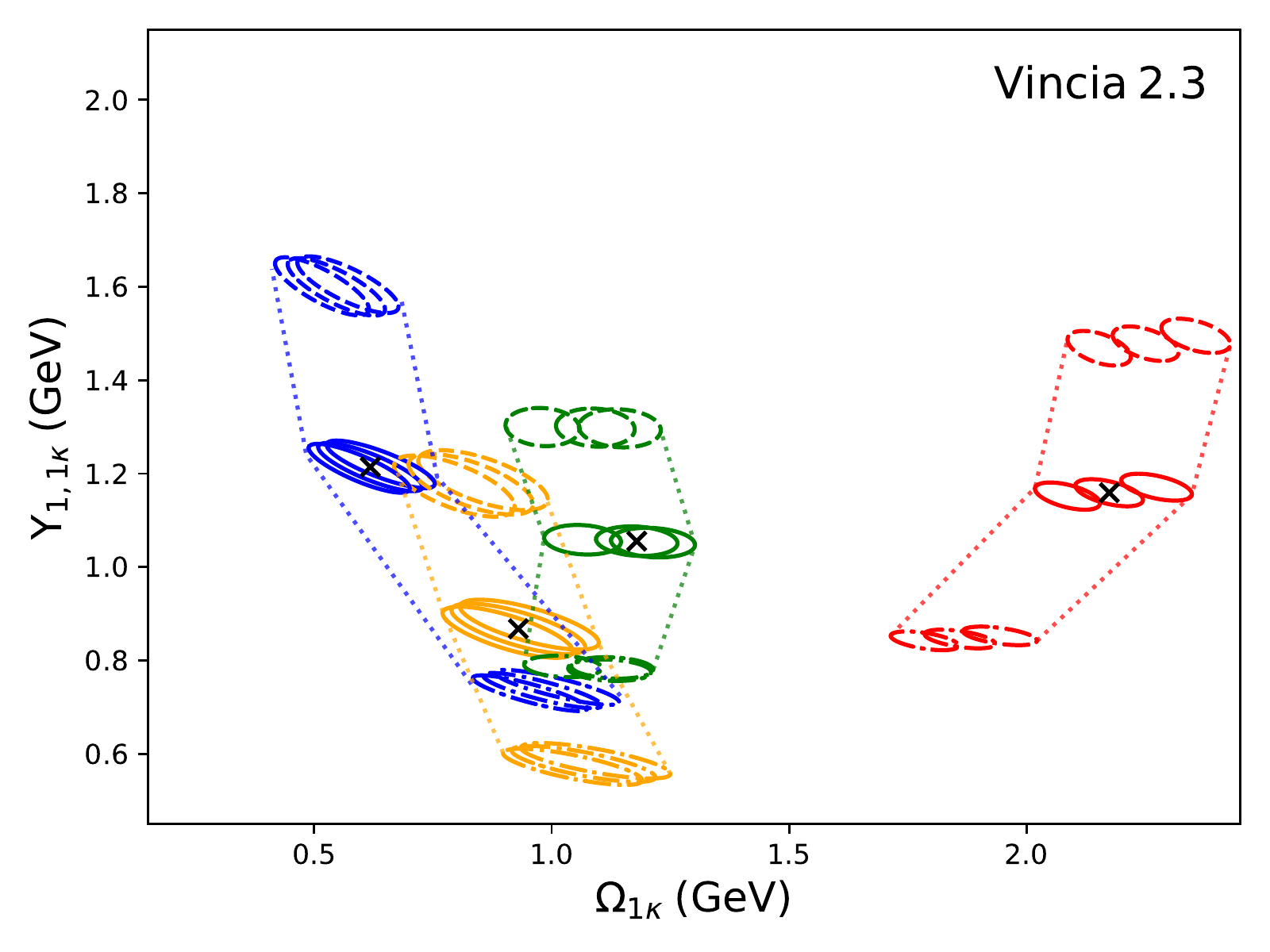}
	\includegraphics[width=0.45\textwidth]{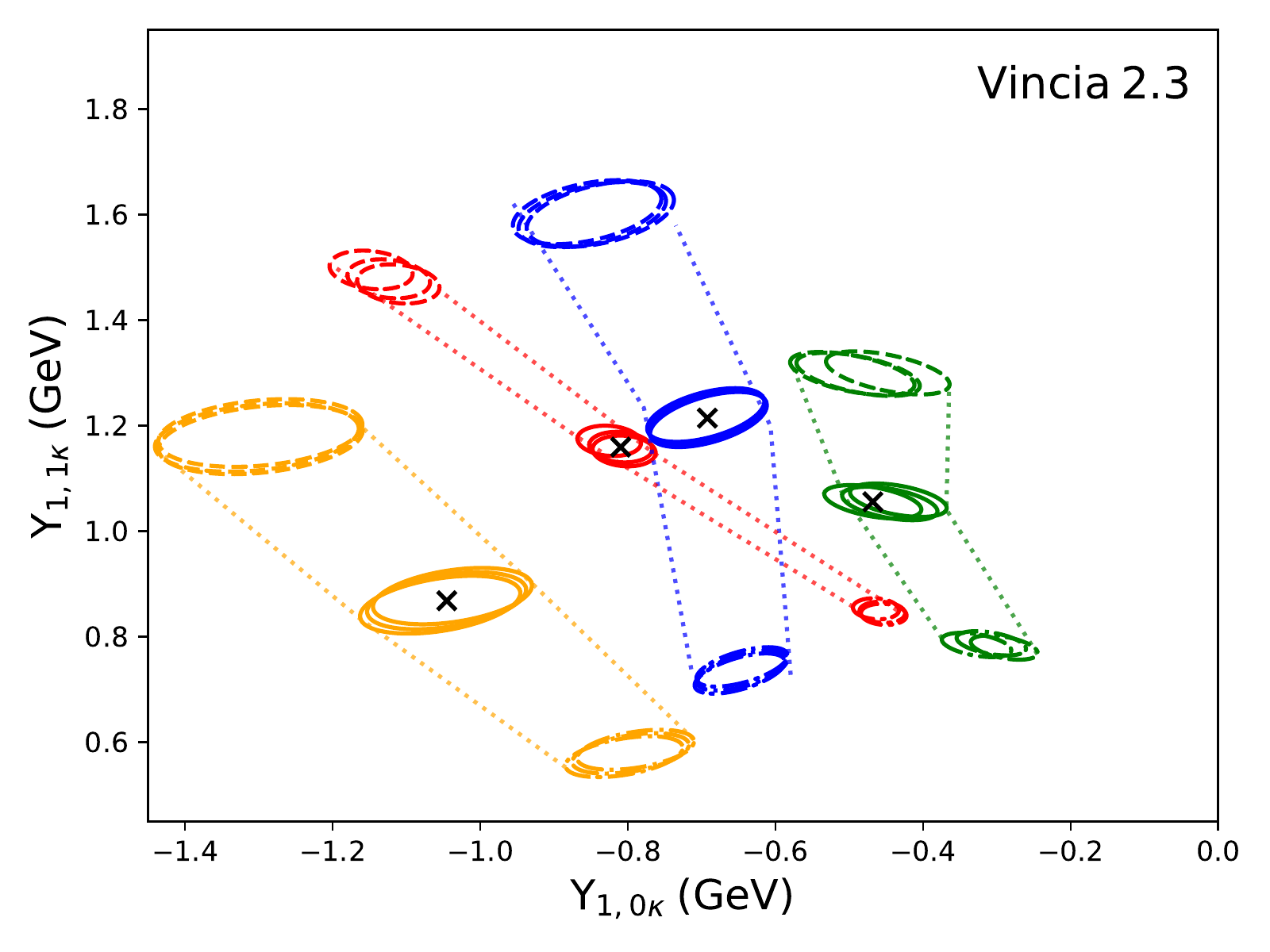}
	\includegraphics[width=0.45\textwidth]{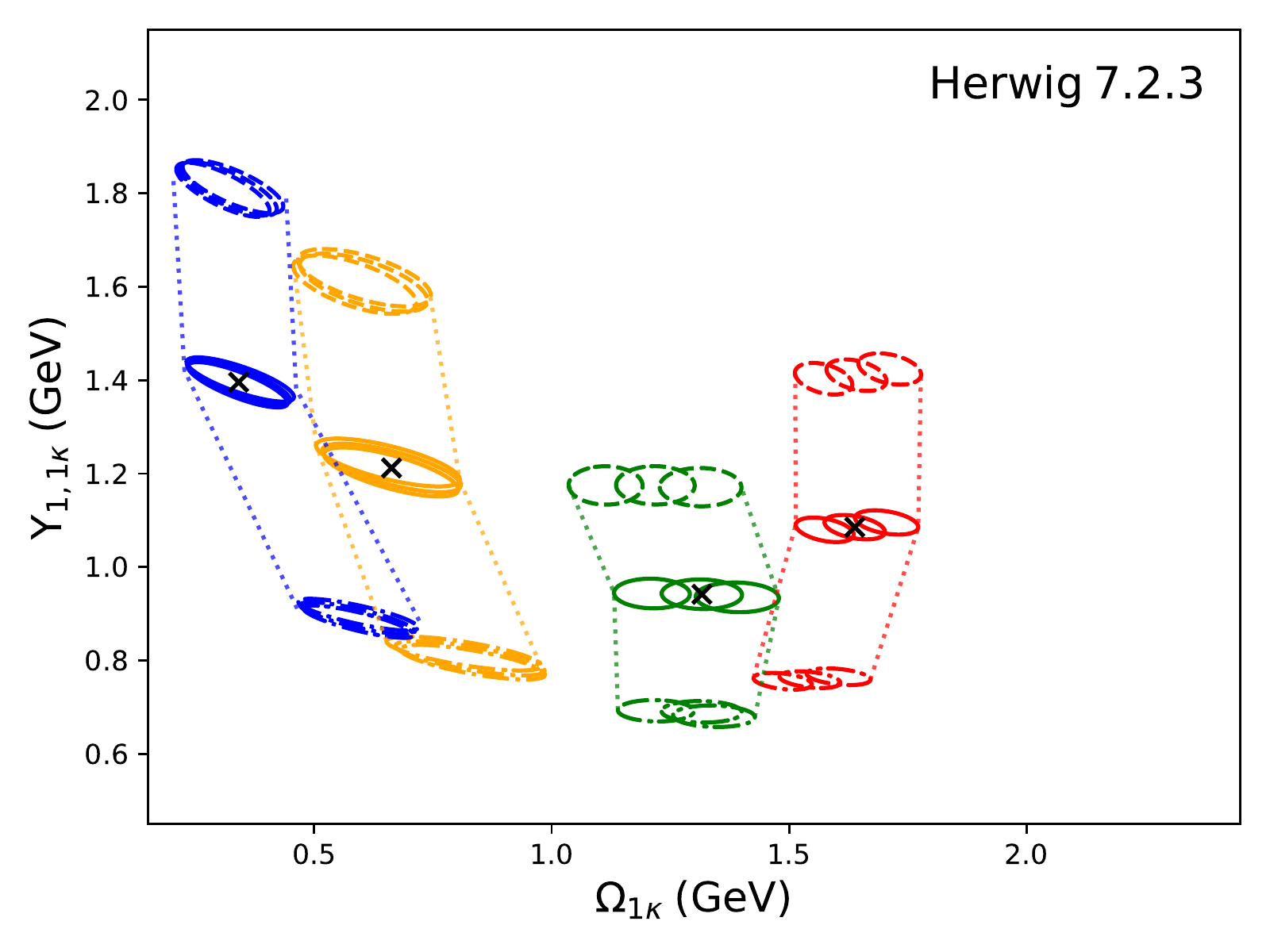}
	\includegraphics[width=0.45\textwidth]{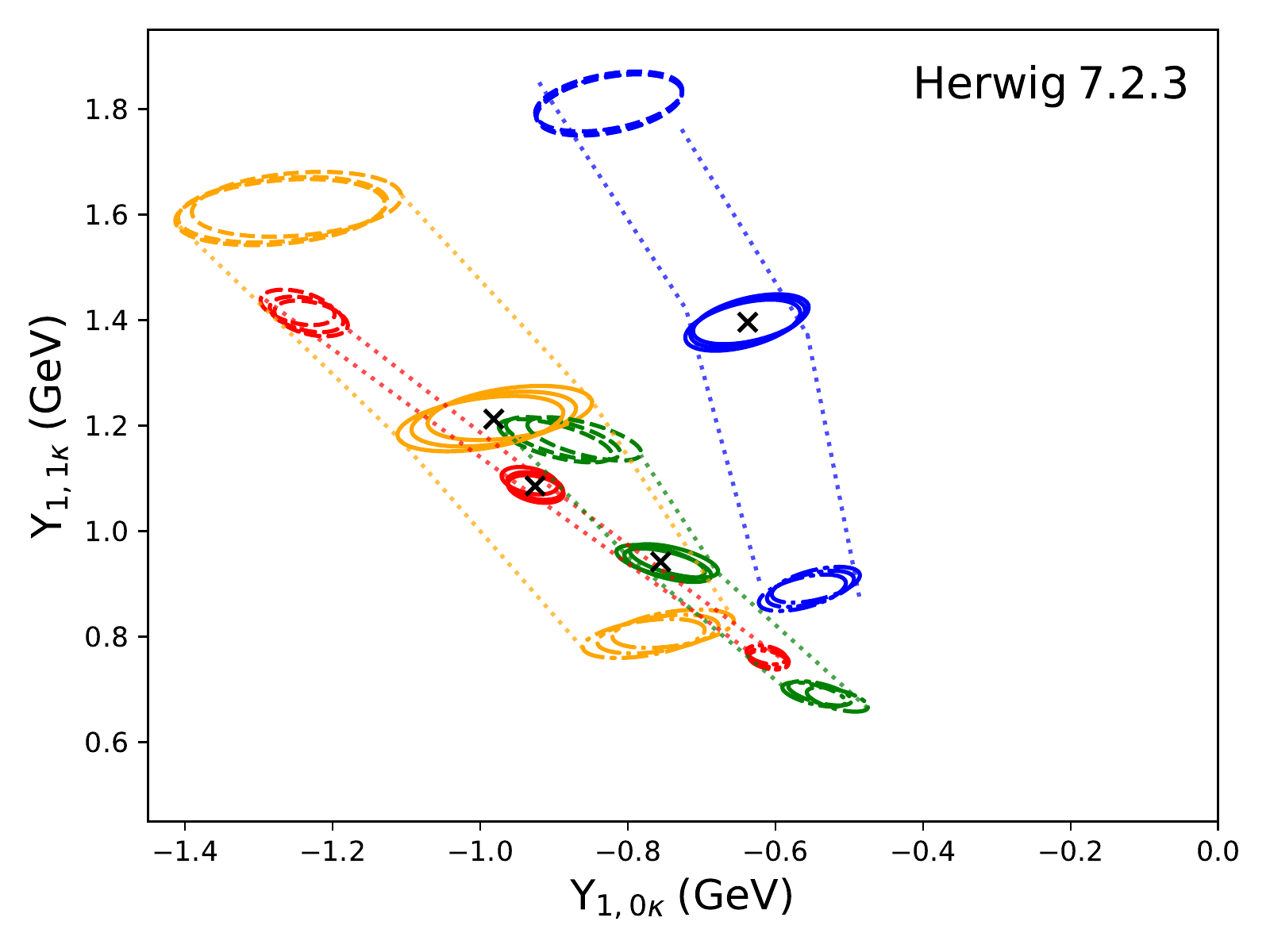}
	\caption{$1\sigma$ contours showing correlations between $\Ukb$ and $\Ok$ (left column) and between $\Ukb$ and $\Uka$ (right column) for \Pythiaxx (top row), \Vinciaxx (middle row) and \Herwigxx (bottom row).}
	\label{fig:contours}
\end{figure}

In order to better visualize the level of agreement for quark and gluon jets in $\ee$ and $pp$ collisions for the three MC event generators considered we show in \fig{contours} the 1$\sigma$ contours for correlations between $\Ukb$ and $\Ok$, and between $\Uka$ and $\Ukb$. The dominant uncertainty results from variation in $C_2^\kappa$. 
Here we see for all three hadronization models a better agreement of correlations between $\Ok$ and $\Uka$ for quark jets in the two colliders than $\Uka$ and $\Ukb$. As noted earlier in the main text, \Pythia and \Vincia results for gluon jets for the two collider settings significantly disagree, whereas the corresponding results for \Herwig display similar trend as quark jets.

\onecolumngrid

	\begin{figure}[t]
		\centering
		\includegraphics[width=0.45\textwidth]{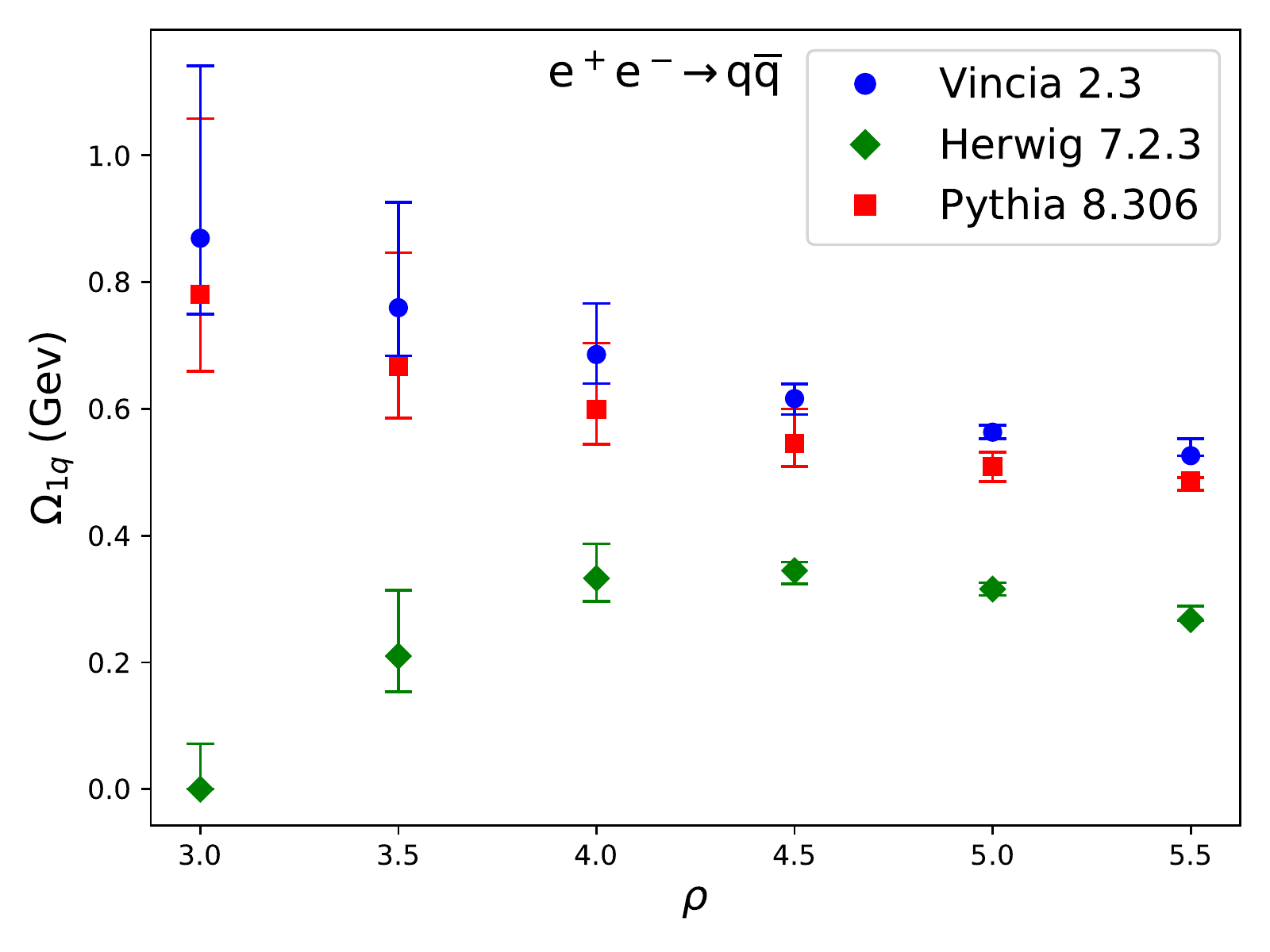}
		\includegraphics[width=0.45\textwidth]{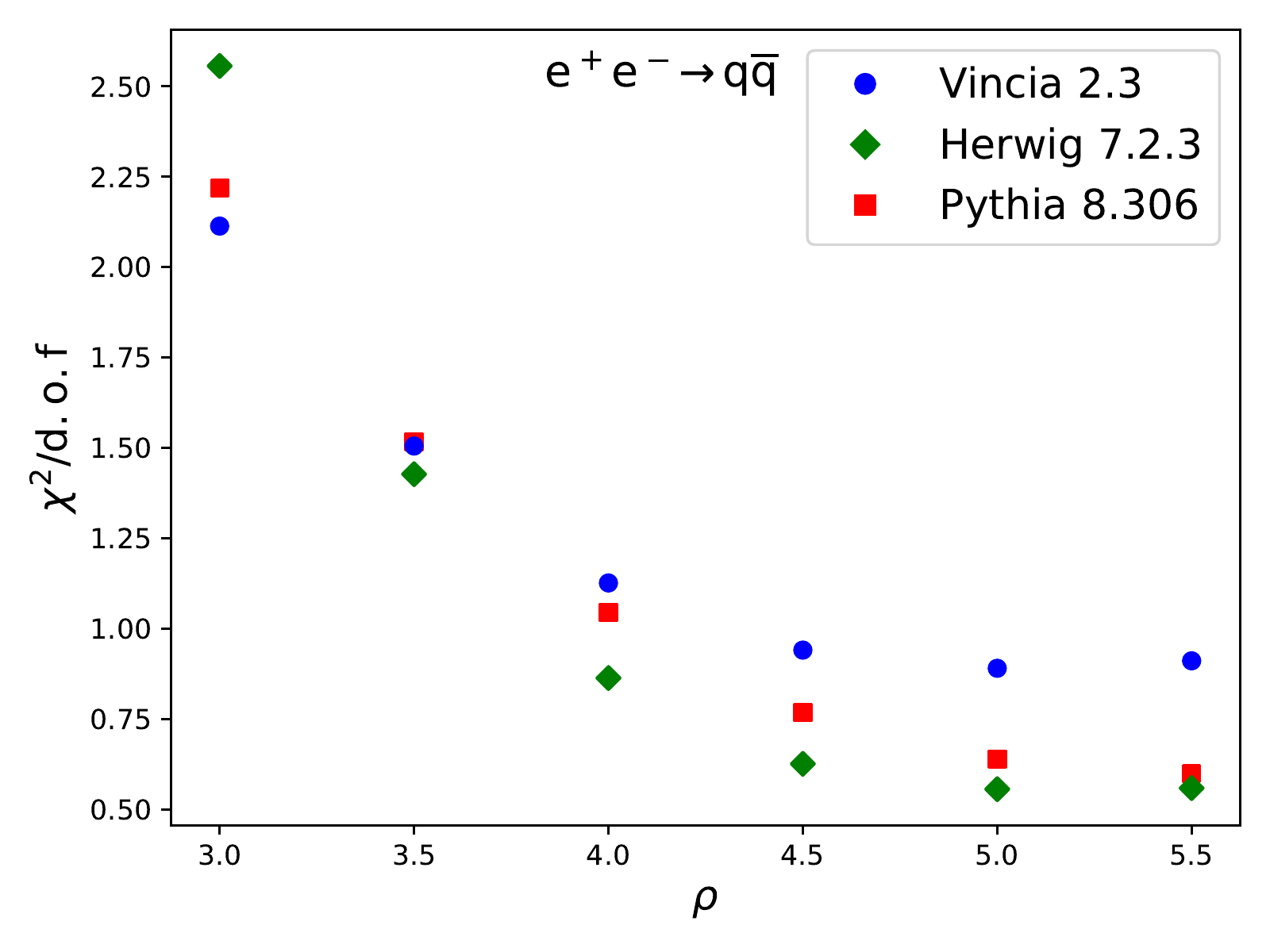}
		\includegraphics[width=0.45\textwidth]{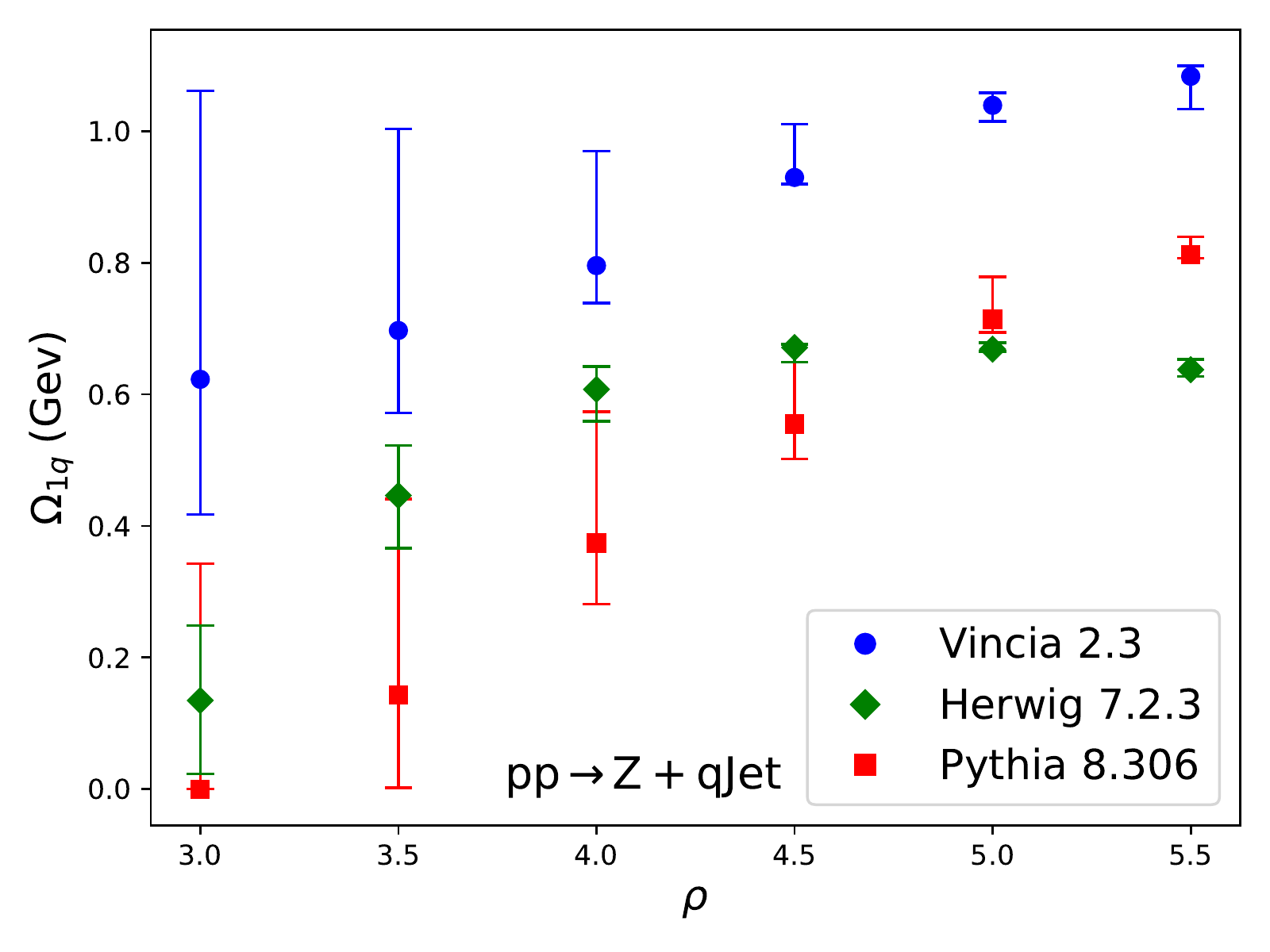}
		\includegraphics[width=0.45\textwidth]{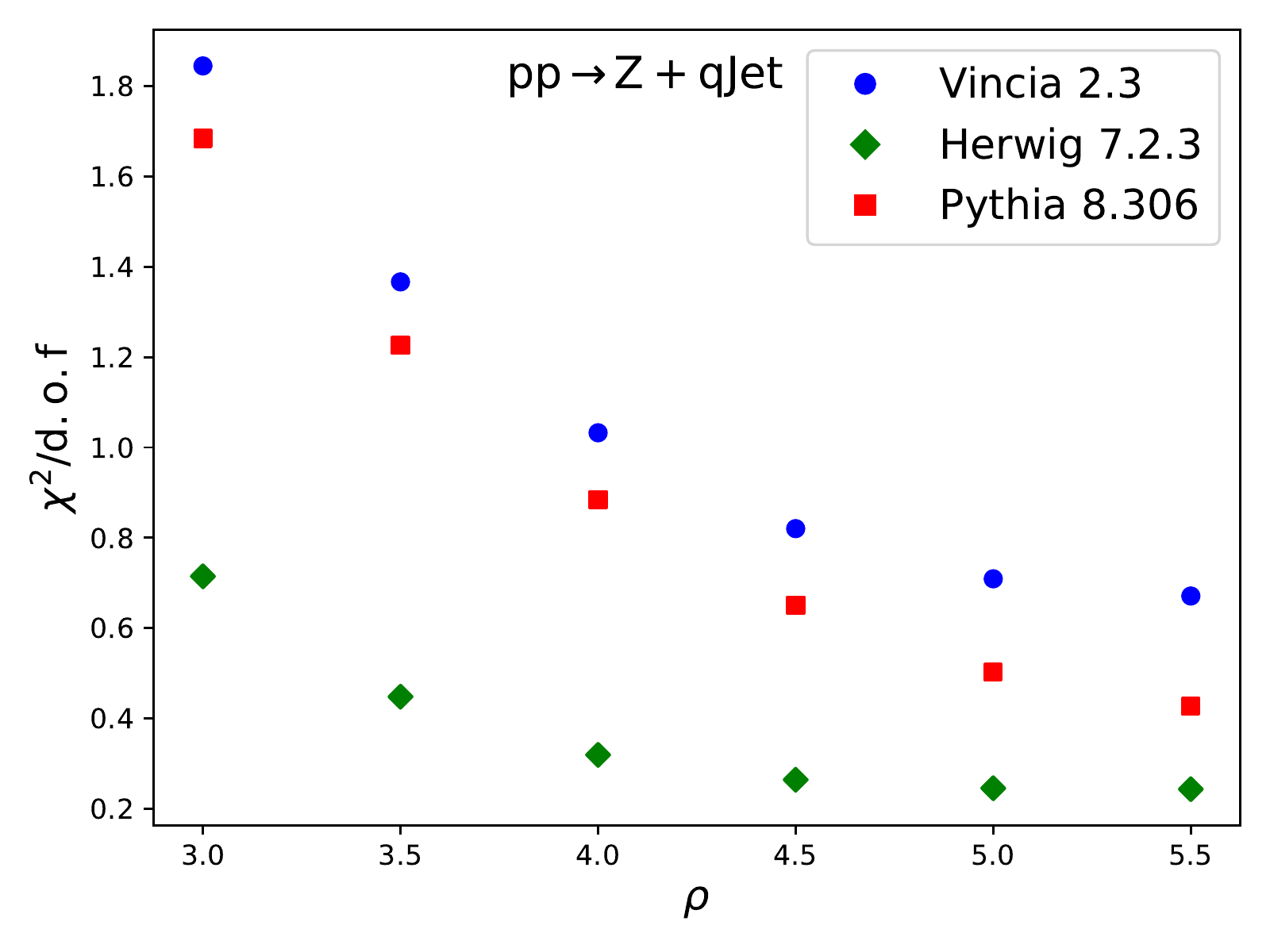}
		\includegraphics[width=0.45\textwidth]{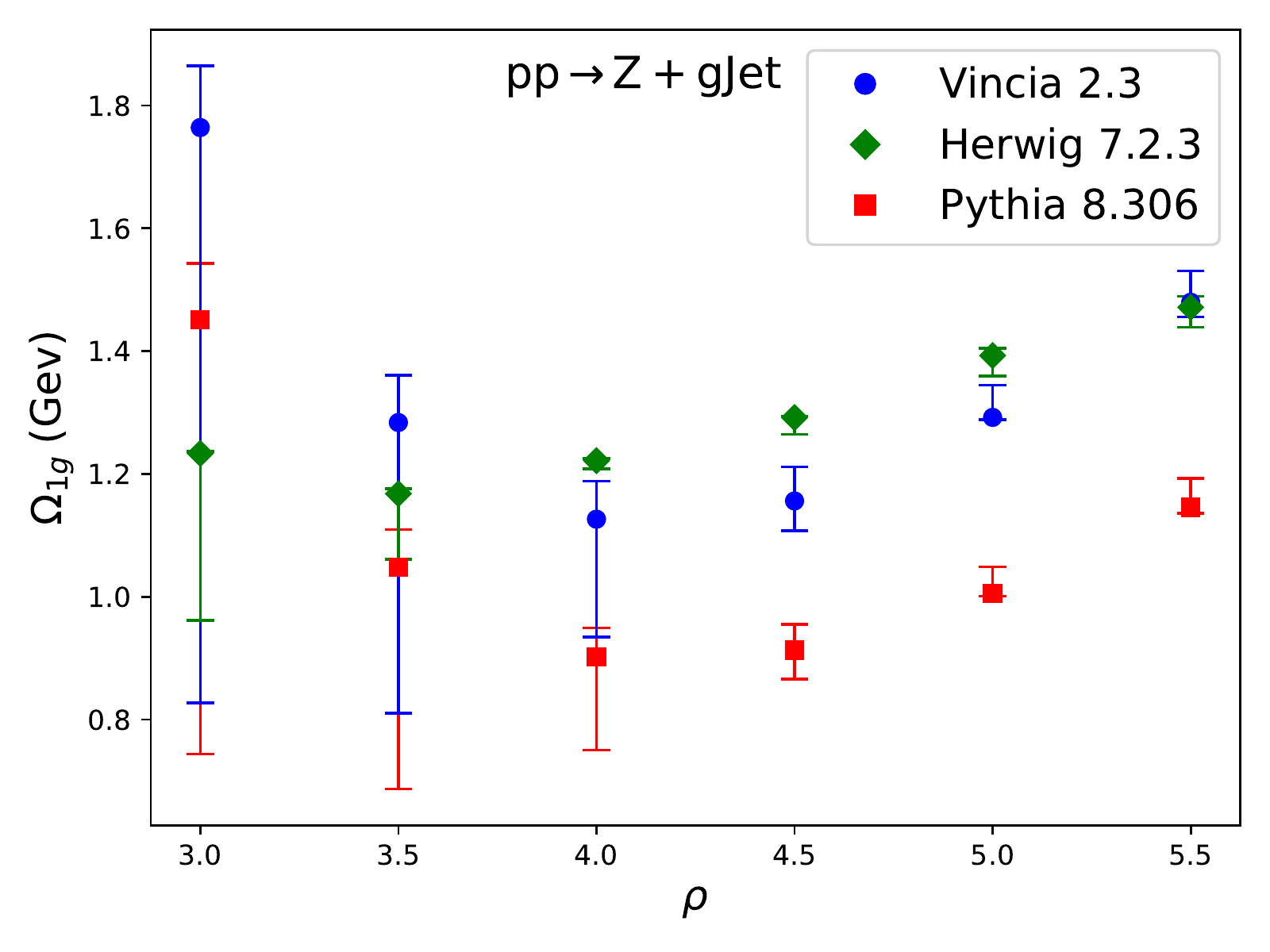}
		\includegraphics[width=0.45\textwidth]{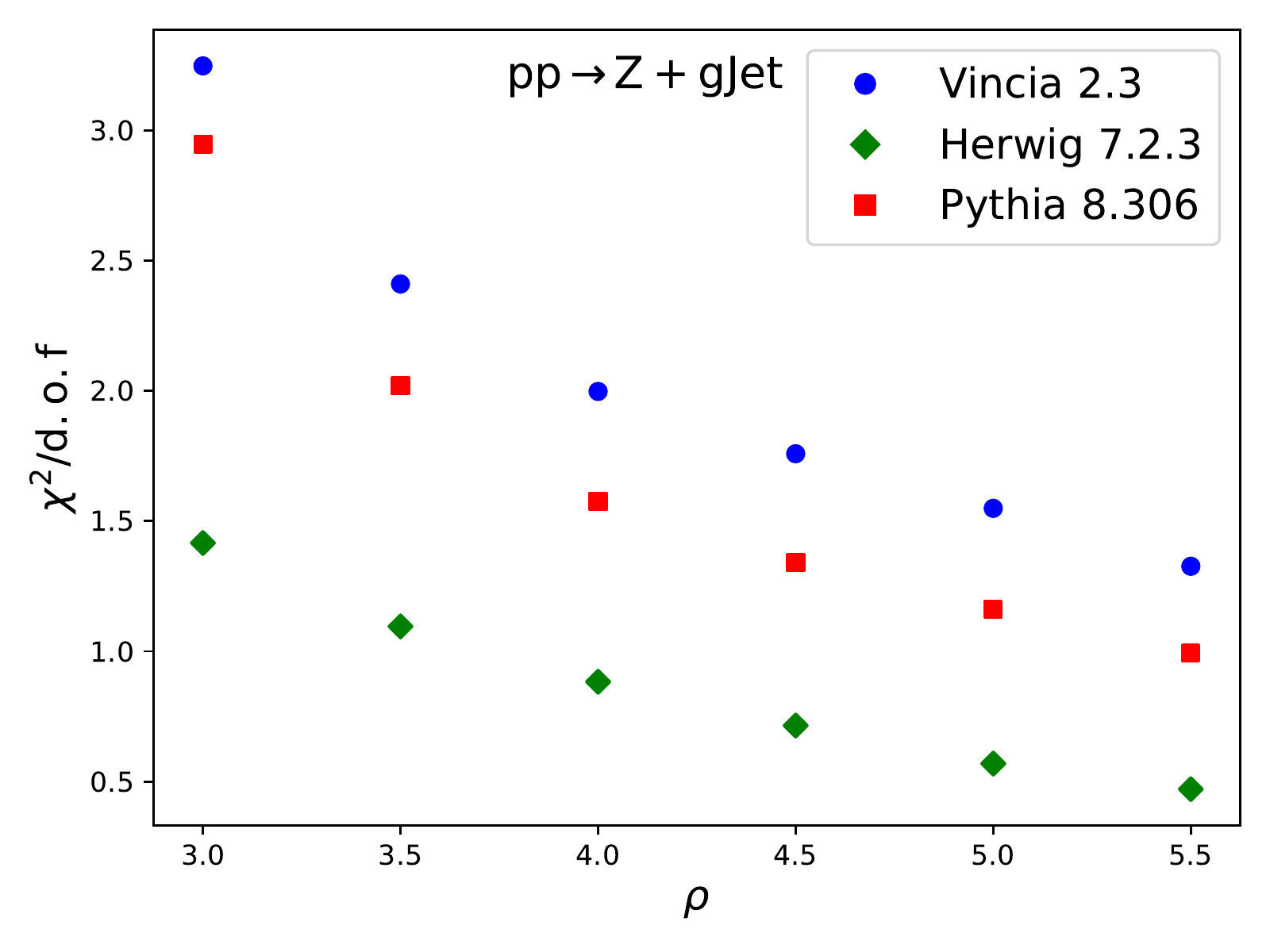}
		\caption{Fit range dependence of the fit values of NP parameters and the reduced $\chi^2$ parameterized in terms of $\rho$ defined in \eq{rhodef}.}
		\label{fig:rho}
	\end{figure}
\newpage
\twocolumngrid

Finally, we now investigate the dependence of the fit values of the NP parameters and the resulting $\chi^2$ on the fit range, parameterized in terms of a parameter $\rho$ defined via the equation
\begin{align}\label{eq:rhodef}
	\xi_{\rm SDOE} \equiv \xi_0 \Big(\frac{\rho \Lambda_{\rm QCD}}{Q \xi_0} \Big)^{\frac{2+\beta}{1+\beta}} \,.
\end{align}
Thus, $\rho$ determines the extent of the SDOE region. In \fig{rho} we illustrate the fit value obtained for $\Ok$ (other parameters not shown for simplicity) and the corresponding reduced $\chi^2$ value for $\ee \ra q\bar q$ and $pp \ra Z + q, g$ jet processes. 
As the value of $\rho$ increases, \eq{rhodef} implies that we move further into the perturbative region, which leads to a smaller errors on the fit value of $\Ok$ and the other two NP parameters due to the reduction in perturbative uncertainty on $C_{1,2}^\kappa$ moments. 
At the same time, increasing $\rho$ also reduces the available fit range which can impact the quality of the fit. 
As a result, we see in \fig{rho} that the reduced $\chi^2$ value initially improves as $\rho$ is increased but eventually saturates as the fit range becomes too small. 
In order to find an optimal balance between perturbative uncertainty and fit range, we used the $\ee \ra q\bar q$ process as a benchmark due to the absence of ISR effects. We chose $\rho = 4.5$ as in all the results presented above, it is at this point the errors and $\chi^2$ values stabilize while still allowing for a reasonable fit range. We also see this trend reflected in the values of $\Ok$ for the $\ee\ra q \bar q$ case. In the $pp$ cases, however, we observe a linear dependence of $\Ok$ on the fit range, which is within the uncertainty for quark jets but not for gluon jets.